\documentclass[11pt,a4paper,preprintnumbers]{article}

\usepackage{jheppub}
\usepackage{graphicx}
\usepackage{amsmath}
\usepackage{xspace}
\usepackage{xcolor}
\usepackage{float}
\usepackage{comment}
\usepackage{subcaption}
\usepackage{textcomp}
\usepackage{placeins}
\usepackage{bm}
\usepackage{pstricks}
\usepackage{color}
\usepackage{braket}
\usepackage{slashed}
\usepackage{soul}

\newcommand{\ord}[1]{\mathcal{O}(#1)}

\newcommand{\rescaleoneplot}{0.49\columnwidth}
\newcommand{\rescaletwoplots}{0.49\textwidth}

\newcommand{\rescalethreeplots}{0.32\textwidth}

\newcommand{\spaceabovefigurecaption}{\vspace*{-2ex}}
\newcommand{\spacebelowfigurecaption}{\vspace*{0ex}}
\newcommand{\df}{\mathrm{d}}

\newcommand{\Tau}{\mathcal{T}}

\newcommand{\GeV}{\,\mathrm{GeV}}

\newcommand{\nn}{\nonumber}

\newcommand{\cP}{\mathcal{P}}

\newcommand{\cut}{\mathrm{cut}}

\newcommand{\NLO}{\mathrm{NLO}}
\newcommand{\NNLO}{\mathrm{NNLO}}

\newcommand{\NS}{\mathrm{NS}}

\newcommand{\zero}{{0}}
\newcommand{\one}{{1}}

\newcommand{\lqcd}{\Lambda_\mathrm{QCD}}
\newcommand{\dsigMC}{\df\sigma^\textsc{mc}}

\newcommand{\de}{\mathrm{d}}

\newcommand{\geneva}{\textsc{Geneva}\xspace}
\newcommand{\dire}{\textsc{Dire}\xspace}
\newcommand{\sherpa}{\textsc{Sherpa}\xspace}
\newcommand{\Matrix}{\textsc{Matrix}\xspace}

\newcommand{\pythia}{\textsc{Pythia}\xspace}

\def\TauZcutre{\ensuremath{\Tau_{0,\rm re}^{\cut}}\xspace}
\def\TauZcutns{\ensuremath{\Tau_{0,\rm ns}^{\cut}}\xspace}
\newcommand{\sss}{\mathchoice %
  {\displaystyle} %
  {\displaystyle} %
  {\scriptscriptstyle} %
  {\scriptscriptstyle} %
}
\newcommand{\MAX}{{\sss\rm max}} % maximum
\newcommand{\MIN}{{\sss\rm min}} % minimum

\newcommand{\DS}{\displaystyle} % fix formulae in arrays and fractions
\renewcommand{\L}{\left}
  \newcommand{\R}{\right}

\renewcommand{\thesection}{\arabic{section}}

\allowdisplaybreaks[2]

\renewcommand{\>}{\,} % +3mu
\newcommand{\<}{\!} % -3mu
\begin{document}

%%%%%%%%%%%%%%%%%%%%%%%%%%%%%%%%%%%%%%%%%%%%%%%%%%%%%%%%%%%%%%%%%%%%%%%%%%%%%%%%
% Title page
%%%%%%%%%%%%%%%%%%%%%%%%%%%%%%%%%%%%%%%%%%%%%%%%%%%%%%%%%%%%%%%%%%%%%%%%%%%%%%%%

\title{Double Higgs production at NNLO interfaced to parton showers in GENEVA}
\preprint{\vbox{\hbox{DESY 22-204}\hbox{UWThPh-2022-19}}}

\author[a]{Simone Alioli,}

\author[a]{Georgios Billis,}

\author[a,b]{Alessandro Broggio,}

\author[a,c]{Alessandro Gavardi,}

\author[a]{Stefan Kallweit,}

\author[c]{Matthew A.~Lim,}

\author[a]{Giulia Marinelli,}

\author[a]{Riccardo Nagar}

\author[a]{and Davide Napoletano}

\affiliation[a]{Universit\`{a} degli Studi di Milano-Bicocca \& INFN, Piazza della Scienza 3, Milano 20126, Italy\vspace{0.5ex}}
\affiliation[b]{Faculty of Physics, University of Vienna, Boltzmanngasse 5, A-1090 Wien, Austria\vspace{0.5ex}}
\affiliation[c]{Deutsches Elektronen-Synchrotron DESY, Notkestr. 85, 22607 Hamburg, Germany\vspace{0.5ex}}
\emailAdd{simone.alioli@unimib.it}
\emailAdd{georgios.billis@unimib.it}
\emailAdd{alessandro.broggio@univie.ac.at}
\emailAdd{alessandro.gavardi@desy.de}
\emailAdd{stefan.kallweit@unimib.it}
\emailAdd{matthew.lim@desy.de}
\emailAdd{g.marinelli10@campus.unimib.it}
\emailAdd{riccardo.nagar@unimib.it}
\emailAdd{davide.napoletano@unimib.it}

\date{\today}

%%%%%%%%%%%%%%%%%%%%%%%%%%%%%%%%%%%%%%%%%%%%%%%%%%%%%%%%%%%%%%%%%%%%%%%%%%%%%%%%
\abstract{ In this work, we study the production
  of Higgs boson pairs at next-to-next-to-leading order in QCD matched to parton
  showers, using the \textsc{Geneva} framework and working in  the heavy-top-limit approximation.
  This includes the resummation of
  large logarithms of the zero-jettiness $\mathcal{T}_0$ up to the next-to-next-to-next-to-leading-log accuracy.
  This process features an extremely large momentum transfer, which makes its study particularly relevant for matching
  schemes such as that employed in \textsc{Geneva}, where the resummation of a
  variable different from that used in the ordering of the parton shower is
  used. To further study this effect, we extend the original shower interface designed  for \textsc{Pythia8}
  to include other parton showers, such as \textsc{Dire} and  \textsc{Sherpa}.
}
%%%%%%%%%%%%%%%%%%%%%%%%%%%%%%%%%%%%%%%%%%%%%%%%%%%%%%%%%%%%%%%%%%%%%%%%%%%%%%%%

\maketitle
\flushbottom
%% optional: table of contents
% \tableofcontents

%%%%%%%%%%%%%%%%%%%%%%%%%%%%%%%%%%%%%%%%%%%%%%%%%%%%%%%%%%%%%%%%%%%%%%%%%%%%%%%%
\section{Introduction}
\label{sec:intro}
%%%%%%%%%%%%%%%%%%%%%%%%%%%%%%%%%%%%%%%%%%%%%%%%%%%%%%%%%%%%%%%%%%%%%%%%%%%%%%%%

After the tenth anniversary of the Higgs boson discovery at the
Large Hadron Collider (LHC)~\cite{ATLAS:2012yve,CMS:2012qbp}, the study of its
fundamental properties is still a very active field of research. Among all of the properties already
measured -- see for example Refs.~\cite{ATLAS:2022tnm,CMS:2022uhn}
and references therein --
or in the processes to be measured~\cite{CMS:2022wqf,ATLAS:2022ers},
one of the most important
parameters connected with the electroweak symmetry breaking (EWSB) mechanism is the
Higgs boson self-coupling, which, so far, has not yet been directly measured~\cite{CMS:2022hgz,CMS:2022kdx,ATLAS:2022xzm,ATLAS:2022hwc}.
Even with the future high-luminosity phase of the LHC, the Higgs
self-coupling is only predicted to be constrained by around 50\%~\cite{DiMicco:2019ngk}.

The measurement of the Higgs self-coupling
is extremely dependent on precise theoretical predictions, and, at hadron
colliders, it can be probed primarily through the production of a pair of Higgs
bosons. Other indirect ways of estimating the self-coupling that
exploit electroweak corrections in high precision observables have been
proposed and studied~\cite{Degrassi:2016wml,Degrassi:2017ucl,Maltoni:2017ims}.
Similar to the production
of a single Higgs boson at hadron colliders, the main production mechanism of a Higgs pair
is through a top-quark loop produced via gluon-gluon fusion~\cite{Glover:1987nx}.
This means that already at the first non-trivial order in perturbation theory,
one has to deal with a one-loop calculation~\cite{Glover:1987nx,Eboli:1987dy,Plehn:1996wb}, which makes the inclusion
of higher-order terms particularly difficult. %~\cite{Borowka:2016ehy,Baglio:2018lrj}.
However, just as in the single
Higgs boson case, one can treat the top quark as being infinitely heavy with
respect to the Higgs boson, thus obtaining an effective gluon-gluon-Higgs vertex~\cite{Glover:1987nx}.
Within this so-called heavy-top limit~(HTL),
the first results at next-to-leading-order~(NLO) accuracy in QCD have been computed in Ref.~\cite{Dawson:1998py}, those at next-to-next-to-leading order~(NNLO) in
Refs.~\cite{deFlorian:2013jea,deFlorian:2016uhr}, and at
next-to-next-to-next-to-leading order (N$^3$LO) in Refs.~\cite{Chen:2019lzz,Banerjee:2018lfq}.
%In the same limit, threshold resummation approximations at next-to-next-leading-logarithmic accuracy~(NNLL)
%matched to NLO have been studied in Ref.~\cite{Shao:2013bz}, and matched to NNLO in~Ref.~\cite{deFlorian:2015moa}.
Beyond the HTL, NLO results with full top-quark mass dependence have been presented in
Refs.~\cite{Borowka:2016ehy,Borowka:2016ypz,Baglio:2018lrj,Baglio:2020ini}. Approximations
to include finite top-quark mass effects in fully differential calculations were discussed in
Refs.~\cite{Grazzini:2018bsd,DeFlorian:2018eng} up to the NNLO and in Ref.~\cite{Chen:2019fhs} at N$^3$LO.
Furthermore, the exact NLO results were combined with a resummation of logarithms of the transverse
momentum of the Higgs pair system to next-to-leading logarithmic~(NLL)
accuracy~\cite{Ferrera:2016prr}, and
a soft-gluon resummation to NLL was performed in Ref.~\cite{DeFlorian:2018eng}.
Exact NLO results matched to the parton shower appeared in
Refs.~\cite{Heinrich:2017kxx,Jones:2017giv,Heinrich:2019bkc}, while techniques
to systematically include finite-mass effects had been studied~\cite{Frederix:2014hta,Maltoni:2014eza,Maierhofer:2013sha} before the exact NLO results became available.
%Refs.~\cite{Grigo:2013rya,Grigo:2014jma,Grigo:2015dia,Frederix:2014hta,Maltoni:2014eza,Maierhofer:2013sha,Degrassi:2016vss,Bonciani:2018omm,Davies:2019djw,Davies:2018qvx,Davies:2019esq,Mishima:2018olh,Bellafronte:2022jmo}.
Several approaches towards analytical results for the two-loop amplitudes with full top-mass dependence based on different expansions were discussed in the literature~\cite{Grigo:2013rya,Grigo:2014jma,Grigo:2015dia,Degrassi:2016vss,Grober:2017uho,Bonciani:2018omm,Davies:2018qvx,Mishima:2018olh,Davies:2019dfy,Davies:2019djw,Davies:2021kex,Bellafronte:2022jmo}.
%Davies:2019esq -> proceeding

In this work we provide the first implementation of the production of a Higgs pair at NNLO QCD
in gluon-gluon fusion, using the HTL, matched to the parton shower. We do so using the
well-established \geneva framework, which has been extensively exploited to provide
fully differential results for various other colour singlet production
processes~\cite{Alioli:2015toa,Alioli:2019qzz,Alioli:2020fzf,Alioli:2020qrd,Alioli:2021egp,Alioli:2021qbf}.
Our results feature the resummation of
the zero-jettiness ($\mathcal{T}_0$) at next-to-next-to-leading-logarithmic accuracy
within the primed counting (NNLL$^\prime$).
While it is well known that the heavy-top approximation does not work as well for double
Higgs production as it does for single
Higgs boson~\cite{Borowka:2016ehy,Jones:2017giv,Grazzini:2018bsd}, the problem of
including finite top-quark (and bottom-quark) mass effects
is largely orthogonal to the problem of matching a NNLO calculation for $gg\to HH$ to the NNLL$^\prime$ resummation of $\Tau_0$
and to the parton shower. For this reason, in this work we neglect all power-like heavy-quark
mass effects, but we note
that for an accurate event generator these would have
to be included. We leave
the study of the inclusion of mass effects to a future publication.
Note that other methods for the matching of NNLO calculations to the parton shower
(NNLOPS in short) are available~\cite{Hoche:2014dla,Monni:2019whf,Campbell:2021svd},
but as far as we are aware no predictions
for this specific process are available to date, using any of these methods.

The outline of this work is the following. First, in Sec.~\ref{sec:theory}, we review the main features of the
\geneva method, highlighting the differences with previous implementations  which have been specifically designed for this process.
Second, in Sec.~\ref{sec:validation}, we validate our results by comparing
to a fixed-order NNLO calculation provided by an independent code,
\Matrix~\cite{Grazzini:2017mhc,deFlorian:2016uhr}.
Next, in Sec.~\ref{sec:shower},  we discuss the impact
of the parton shower, presenting our results. Finally, in
Sec.~\ref{sec:conclusions} we present our conclusions.

%%%%%%%%%%%%%%%%%%%%%%%%%%%%%%%%%%%%%%%%%%%%%%%%%%%%%%%%%%%%%%%%%%%%%%%%%%%%%%%%
\section{Theoretical framework}
\label{sec:theory}
\begin{figure}[t]
  % \begin{center}
  %   \begin{tikzpicture}[line width=0.9 pt, scale=1.1]
  %     \draw[psgluon,opacity=1.]        (-1.0,1.0)    -- (0.0,0.) node[above=3pt,pos=0.3] {$g$};
  %     \draw[psgluon,opacity=1.]        (-1.0,-1.0)   -- (0.0,0.) node[below=3pt,pos=0.3] {$g$};
  %     \fill[color=black,draw] (0.,0.) circle [radius=0.15];
  %     \draw[scalar,opacity=1.]         (0.,0.0)     -- (1.,0);
  %     \draw[scalar,opacity=1.]         (1.,0.0)     -- (2.,1)    node[above=3pt,pos=0.7] {$H$};
  %     \draw[scalar,opacity=1.]         (1.,0.0)     -- (2.,-1)   node[below=3pt,pos=0.7] {$H$};
  %   \end{tikzpicture}\hspace{100pt}
  %   \begin{tikzpicture}[line width=0.9 pt, scale=1.1]
  %     \draw[psgluon,opacity=1.]        (-1.0,1.0)    -- (0.5,0.) node[above=3pt,pos=0.3] {$g$};
  %     \draw[psgluon,opacity=1.]        (-1.0,-1.0)   -- (0.5,0.) node[below=3pt,pos=0.3] {$g$};
  %     \fill[color=black,draw] (0.5,0.) circle [radius=0.15];
  %     \draw[scalar,opacity=1.]         (0.5,0.0)     -- (2.,1)    node[above=3pt,pos=0.7] {$H$};
  %     \draw[scalar,opacity=1.]         (0.5,0.0)     -- (2.,-1)   node[below=3pt,pos=0.7] {$H$};
  %   \end{tikzpicture}
  % \end{center}
% \begin{figure}[tp]
  \begin{center}
    \includegraphics[width=0.35\textwidth]{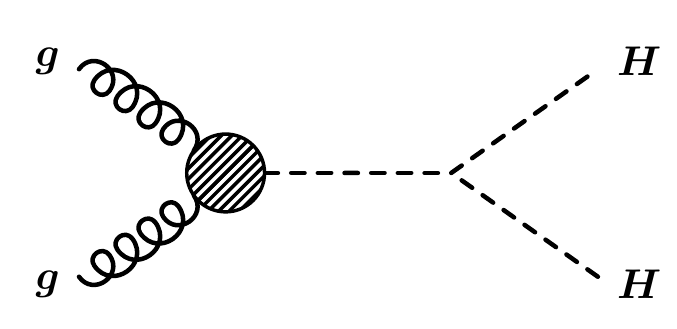}\hspace{20pt}
    \includegraphics[width=0.35\textwidth]{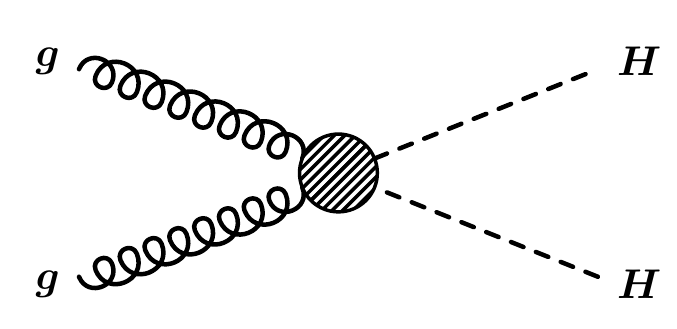}
  \end{center}
  \spaceabovefigurecaption
  \caption{  \label{fig:diagrams} Leading-order diagrams for the production of a Higgs pair
    in the HTL.}
  \spacebelowfigurecaption
\end{figure}

%%%%%%%%%%%%%%%%%%%%%%%%%%%%%%%%%%%%%%%%%%%%%%%%%%%%%%%%%%%%%%%%%%%%%%%%%%%%%%%%
% - process definition, which diagrams we include and approximations used HTL\\
% - justify HTL and discuss mass effects (and why we can forget about them for
% now)
At hadron colliders, the production of a pair of Higgs bosons proceeds via two
mechanisms. In the Standard Model (SM), at tree level, two Higgs bosons can be produced through heavy-quark (charm
or bottom) annihilation through $s$- and $t$-channel diagrams~\cite{Dawson:2006dm}. At
one-loop, one can instead produce a pair of Higgs bosons via a top-quark loop in
triangle and box topologies in the gluon-gluon fusion channel. In the SM,
the latter represents the dominant production mode,
despite being suppressed by two powers of $\alpha_S$
compared to
heavy-quark annihilation. This is due to the larger values of the
gluon parton distribution function (PDF) at the relevant momentum transfer,
combined with the larger coupling of the top quark to the Higgs boson.
In the same spirit as the production of a single Higgs boson, to simplify
the inclusion of higher-order
effects, one can make the approximation that the top-quark mass is much larger
than the hard scale of the process. This produces effective vertices directly coupling Higgs bosons and (two) gluons, and is normally referred to as HTL. We show
in Fig.~\ref{fig:diagrams} the two leading-order~(LO) diagrams contributing to this
process in this approximation.

It is important to stress that
it is well known that the HTL is a much worse approximation in this particular
case~\cite{Borowka:2016ehy,Jones:2017giv,Grazzini:2018bsd} compared to
single Higgs boson production.
There are two main causes for this. First, as this approximation receives corrections
of the order $\mathcal{O}\left(Q^2/m_t^2\right)$, taking as a typical value
for
the momentum transfer $Q$ the peak of the invariant-mass
distribution of the Higgs pair instead of the Higgs boson mass, we see that
corrections are roughly three to four times larger for di-Higgs than for single
Higgs production\footnote{Note that this is by itself a poor
approximation, as the invariant-mass spectrum has a wide distribution, well sustained up to very large values of $M_{HH}$, meaning that the average invariant-mass value is
actually much larger than the position of the peak, and our previous reasoning only provides an underestimate of the actual corrections.}. Second, the two
diagrams depicted in Fig.~\ref{fig:diagrams} arise
from two different loop topologies: the triangle (left) and the box (right).
While the latter is relatively well approximated in the HTL, the former is
known, from the single scalar production case, to be a poor approximation for
larger invariant masses of
the $s$-channel particle. In this case, this is the mass of the virtual Higgs boson which can
acquire extremely large values, thus spoiling the approximation.
Moreover, in the exact result it is known that the interference between these two
diagrams is large and negative, thus causing a large overall
cancellation  -- this is not well-captured in the HTL.

In order to have a realistic event generator for this process,
heavy-quark mass effects have to be included, and ways to include them order-by-order using
approximants have been studied~\cite{Davies:2019nhm}. However, the present work is mainly concerned
with the effects of matching the NNLO calculation for the FO production
to the resummation of $\mathcal{T}_0$ and the subsequent matching to the parton
shower, for this process. For these aspects, mass effects can be safely considered an orthogonal
problem as they are not expected to considerably change the conclusions drawn in
this paper. Lastly,
due to the lack of available experimental data, for the time being we can safely ignore these effects.
In the same spirit, when discussing the interface to the parton shower
in Sec.~\ref{sec:shower}, we also ignore
all effects given by the hadronisation and fragmentation of hadrons in the
shower, as well as those coming from multi-parton scattering.

In the rest of this section we recap the main features of the \geneva method,
highlighting the main differences with respect to quark-induced processes that
have been considered so far, and stressing the novelties developed during the
completion of this work.

%%%%%%%%%%%%%%%%%%%%%%%%%%%%%%%%%%%%%%%%%%%%%%%%%%%%%%%%%%%%%%%%%%%%%%%%%%%%%%%%
\subsection{The \geneva method}
\label{sec:geneva}
%%%%%%%%%%%%%%%%%%%%%%%%%%%%%%%%%%%%%%%%%%%%%%%%%%%%%%%%%%%%%%%%%%%%%%%%%%%%%%%%

Since the \geneva framework has already been
extensively discussed in previous
publications~\cite{Alioli:2015toa,Alioli:2019qzz,Alioli:2020qrd,Alioli:2021egp,
  Alioli:2021qbf}, for the sake of conciseness, we explicitly refrain from entering into
the details of the method and  only briefly recall the general formulae,
highlighting some key features that are important for this
process.

\geneva employs IR-safe resolution variables to discriminate between events,
which are classified as having 0, 1 or 2 jets according to whether the value of
a given resolution variable is smaller or larger than a given cutoff.
Within this framework, from which the definition of ``physical events'' in  a
Monte Carlo generator naturally follows~\cite{Alioli:2013hqa},
unresolved emissions below the resolution cutoff are integrated over, and the IR
safety of each event and at each perturbative order is ensured.
For all the processes previously studied, a single cut in the
resolution variable discriminating between zero or more jets, either on
zero-jettiness $\Tau_0^\cut$ or on the transverse momentum of the colour singlet
$q_T^\cut$,
had always been used both for
the resummed and the nonsingular components of the calculation.
However, it is perfectly
legitimate to move as much of the nonsingular contribution as possible into the
resolved region of the phase space, such that it can be better described with
the full event kinematics, while still maintaining the IR safety
requirements.
We recall that the variable $N$-jettiness~\cite{Stewart:2010tn} is defined as
\begin{align}
 \Tau_N = \sum_k \min \left[ \hat{q}_a \cdot p_k, \hat{q}_b \cdot p_k, \hat{q}_1
  \cdot p_k, ... , \hat{q}_N \cdot p_k \right] \,,
\end{align}
with $N = 0,1$, $\hat{q}_{a,b}$ are the beam directions and $\hat{q}_k$ represents any final-state massless four-vector that minimises $\Tau_N$, and $p_k$ are the momenta of the final-state partons. Therefore, in the following we explicitly
use \zero-jettiness as our main resolution variable and consider two separate
cuts:
\TauZcutre acting on the resummed singular contribution and \TauZcutns acting on the
nonsingular.
These effectively replace the common $\Tau_0^\cut$
used in previous \geneva implementations.
Choosing \TauZcutns smaller than \TauZcutre allows us to push down the
calculation of the nonsingular contributions to lower values, thereby reducing
the subleading power corrections. Note, however, that the result should be
independent of the exact choice of \TauZcutre, modulo higher-order corrections.

With these definitions, the differential cross sections for the production of events
with $0,1$, and  $\geq 2$ emissions are given by\footnote{With N$^k$LO$_l$ we refer to the FO calculation at $k$th order in QCD for the final state with $l$ resolved jets.}
%%%
\begin{align}
  \frac{\dsigMC_0}{\df\Phi_0}(\TauZcutre,\TauZcutns) &= \frac{\df\sigma^{\rm NNLL^\prime}}{\df\Phi_0}(\TauZcutre) - \frac{\df\sigma^{\rm NNLL^\prime}}{\df\Phi_{0}}(\TauZcutns  )\bigg\vert_{\NNLO_0} \nn \label{eq:0full}\\
                                                     &\qquad +(B_0+V_0+W_0)(\Phi_0)\, +\,  \int \frac{\mathrm{d} \Phi_1}{\mathrm{d} \Phi_0} (B_1 + V_1)(\Phi_1)\,\theta\big( \Tau_0(\Phi_1)< \TauZcutns \big) \nn \\
                                                     &\qquad+  \int \frac{\mathrm{d} \Phi_2}{\mathrm{d} \Phi_0} \,B_2 (\Phi_2) \, \theta\big( \Tau_0(\Phi_2)< \TauZcutns\big)\,,
\end{align}

\begin{align}
  \label{eq:1masterfull}
  \frac{\dsigMC_{1}}{\df\Phi_{1}} &(\Tau_0 > \TauZcutns;\TauZcutre; \Tau_{1}^\cut) = \Bigg\{ \frac{\df\sigma^{\rm NNLL^\prime}}{\df\Phi_0\df\Tau_0} \cP(\Phi_1)\, \theta(\Tau_0 > \TauZcutre)   \\
                                  & + \bigg[ (B_1 + V_1^C)(\Phi_1) - \frac{\df\sigma^{\rm NNLL^\prime}}{\df\Phi_0\df\Tau_0}  \bigg\vert_{\NLO_1} \cP(\Phi_1) \bigg]\theta(\Tau_0 > \TauZcutns) \Bigg\} \times\, U_1(\Phi_1, \Tau_1^\cut) \nn \\
                                  &+\int\ \bigg[\frac{\df\Phi_{2}}{\df\Phi^\Tau_1}\,B_{2}(\Phi_2) \, \theta\!\left(\Tau_0(\Phi_2) > \TauZcutns\right)\,\theta(\Tau_{1} < \Tau_1^\cut)  - \frac{\df\Phi_2}{\df \Phi^C_1}\, C_{2}(\Phi_{2}) \, \theta(\Tau_0 > \TauZcutns) \bigg] \nn \\
                                  &- B_1(\Phi_1)\, U_1^{(\one)}(\Phi_1, \Tau_1^\cut)\, \theta(\Tau_0 > \TauZcutns)\,,\nn
\end{align}

\begin{align}
  \label{eq:2masterfull}
  \frac{\dsigMC_{\geq 2}}{\df\Phi_{2}}& (\Tau_0 > \TauZcutns; \TauZcutre; \Tau_{1}>\Tau_{1}^\cut) = \\& \Bigg\{ \Bigg[(B_1 + V_1^C)({\Phi}_1) - \frac{\df\sigma^{\rm NNLL^\prime}}{\df\Phi_0\df\Tau_0}\bigg|_{\NLO_1} \cP({\Phi}_1) \Bigg] \theta(\Tau_0 > \TauZcutns) \Big\vert_{{\Phi}_1 = \Phi_1^\Tau\!(\Phi_2)} \bigg] \nn \\
                                      &+ \frac{\df\sigma^{\rm NNLL^\prime}}{\df\Phi_0\df\Tau_0} \, \cP({\Phi}_1) \, \theta(\Tau_0 > \TauZcutre) \Big\vert_{{\Phi}_1 = \Phi_1^\Tau\!(\Phi_2)} \Bigg\} \,  U_1'({\Phi}_1, \Tau_1) \, \cP(\Phi_2) \, \theta(\Tau_1 > \Tau_1^\cut) \nn  \\
                                      &+  B_2(\Phi_2)\, \theta(\Tau_{1}>\Tau^{\mathrm{cut}}_{1}) \theta\left(\Tau_0(\Phi_2) > \TauZcutns\right)\, \nn \\
                                      &- B_1(\Phi_1^\Tau)\,U_1^{(\one)\prime} \big({\Phi}_1, \Tau_1\big)\,\cP(\Phi_2)\, \Theta(\Tau_1 > \Tau_1^\cut)\, \theta\left(\Tau_0(\Phi_2) > \TauZcutns\right)\,. \nn
\end{align}
In the previous formulae  $B_{0,1,2}$ represents the $0,1,2$-parton tree-level contributions, $V_{0,1}$ the $0,1$-parton one-loop contributions and $W_0$ the two-loop contributions.
We have introduced the notation
\begin{align}
  \frac{\de \Phi_M}{\de \Phi_N^{\mathcal{T}}} = \de \, \Phi_M \delta [\Phi_N - \Phi_N^{\mathcal{T}}(\Phi_M)]\, \Theta^{\mathcal{T}}(\Phi_M), \qquad N \leq M\,,
\end{align}
to indicate that the integration over a region of the $M$-body phase space is
done keeping the $N$-body phase space and the value of the observable $\mathcal{T}$
fixed.
The $\Theta^{\mathcal{T}}(\Phi_N)$ term limits the integration region to
the phase space points included in the singular contribution for the
observable $\cal{T}$.
The $V_1^C$ term includes contributions of soft and collinear origin and it is defined as
\begin{align}
  V_1^C (\Phi_1) = V_1(\Phi_1) + \int \frac{\de \Phi_2}{\de \Phi_1^C} \, C_2(\Phi_2)\,,
\end{align}
where $C_2$ acts as a local NLO subtraction counterterm that reproduces the singular behaviour of $B_2$. The subtraction counterterms are integrated over the radiation variables $\frac{\de \Phi_2}{\de \Phi_1^C}$ considering the singular limit $C$ of the phase space mapping.\\
We have also introduced normalised splitting functions $\mathcal{P}(\Phi_{N+1})$ to make
the resummed $\Tau_N$ spectrum fully differential in $\Phi_{N+1}$.\
These splitting functions are normalised such that
\begin{align}
  \int \mathcal{P}(\Phi_{N+1}) \, \frac{\de \Phi_{N+1}}{\de \Phi_N \, \de \Tau_N} = 1,
\end{align}
where two extra emission variables, the energy ratio $z$ and the
azimuthal angle $\phi$, are needed besides $\Tau_N$ to define a
splitting $\Phi_N \to \Phi_{N+1}$. The normalised
splitting probability is given by
% \begin{eqnarray}
    %     \label{eq:splittingprob}
    %     && \EI P\<\L(\Phi_{N+1}\R) =
    %   %
             %              \nonumber \\
             %   %
    %           && \EI \frac{\DS f_{kj}\<\L(\Phi_N, \Tau_N, z\R)}{\DS
                   %                    \sum_{k'=1}^{N+2} \int_{z_\MIN^{k'}\<\L(\Phi_N,
                   %                    \Tau_N\R)}^{z_\MAX^{k'}\<\L(\Phi_N, \Tau_N\R)} \df z' \>
                   %                    J_{k'}\<\L(\Phi_N, \Tau_N, z'\R) I_\phi^{k'}\<\L(\Phi_N, \Tau_N,
                   %                    z'\R) \sum_{j'=1}^{n_{\rm split}^{k'}} f_{k'j'}\<\L(\Phi_N,
                   %                    \Tau_N, z'\R)},
                   %   %
                   %                    \nonumber \\
    %   \end{eqnarray}
\begin{equation}
  \label{eq:splittingprob}
  \small{
   \mathcal{P}\<\L(\Phi_{N+1}\R) =
    \frac{\DS f_{kj}\<\L(\Phi_N, \Tau_N, z\R)}{\DS
      \sum_{k'=1}^{N+2} \int_{z_\MIN^{k'}\<\L(\Phi_N,
        \Tau_N\R)}^{z_\MAX^{k'}\<\L(\Phi_N, \Tau_N\R)} \df z' \>
      J_{k'}\<\L(\Phi_N, \Tau_N, z'\R) I_\phi^{k'}\<\L(\Phi_N, \Tau_N,
      z'\R) \sum_{j'=1}^{n_{\rm split}^{k'}} f_{k'j'}\<\L(\Phi_N,
      \Tau_N, z'\R)}}\,,
\end{equation}
where
\begin{equation}
  I_\phi^k\<\L(\Phi_N,\Tau_N,z\R) = \phi_\MAX^k\<\L(\Phi_N,\Tau_N,z\R)
  - \phi_\MIN^k\<\L(\Phi_N,\Tau_N,z\R).
\end{equation}
In the formula above $f_{kj}\L(\Phi_N,\Tau_N, z\R)$ are generic functions based on the Altarelli-Parisi splitting functions,
$z_{\MIN,\MAX}(\Phi_N, \Tau_N, z)$, $\phi_{\MIN,\MAX}(\Phi_N, \Tau_N, z)$ are the integration limits, respectively, in $z$ and $\phi$
defined in Ref.~\cite{Gavardi:2023oho},
and $J\L(\Phi_N, \Tau_N, z\R)$ is the Jacobian related to the change
of variable.
Further details about this new implementation of the splitting probabilities are
discussed in a separate publication~\cite{Alioli:2023har}.

In equations \eqref{eq:1masterfull} and \eqref{eq:2masterfull} we introduce the
Sudakov factor $U_1(\Phi_1, \Tau^{\cut}_1)$ which resums the dependence of
$\Tau^{\cut}_1$ to next-to-leading-logarithmic (NLL) accuracy.
In addition to the formulae for the quark channels already presented in
Ref.~\cite{Alioli:2016wqt}, for gluon-initiated processes we have
\begin{align}
  U^{ggg}_1(\Phi_1, \Tau^{\cut}_1) =  \frac{U}{\Gamma \bigg( 1 + 6 \, C_A \, \eta^{\rm NLL}_{\rm cusp}(\mu_S, \mu_H) \bigg)}\,,
\end{align}
with $\Gamma$ the Euler gamma function and
\begin{align}
  \ln U &= 6 \, C_A \bigg[ 2 \, K^{\rm NLL}_{\Gamma_{\rm cusp}}(\mu_J, \mu_H) - K^{\rm NLL}_{\Gamma_{\rm cusp}}(\mu_S, \mu_H) \bigg] \nn \\
        & + C_A \bigg[ - \ln \bigg( \frac{Q^2_a Q^2_b Q^2_J}{\mu^6_H} \bigg) \, \eta^{\rm NLL}_{\Gamma_{\rm cusp}}(\mu_J, \mu_H) + \ln \bigg( \frac{Q^2_a Q^2_b Q^2_J}{stu} \bigg) \, \eta^{\rm NLL}_{\Gamma_{\rm cusp}}(\mu_S, \mu_H) \bigg] \nn \\
        &-6 \, \gamma_E \, C_A \, \eta^{\rm NLL}_{\Gamma_{\rm cusp}}(\mu_S, \mu_J)  +3 \, K^{\rm NLL}_{\gamma^g_J}(\mu_J, \mu_H).
\end{align}
The functions appearing in the formula above are common in the SCET
literature, see e.g.\ Ref.~\cite{Berger:2010xi}, and are given by
\begin{align}
  K^{\rm NLL}_{\Gamma_{\rm cusp}}(\mu_1, \mu_2) &= - \frac{\Gamma_0}{4 \beta^2_0} \bigg[\frac{4 \pi}{\alpha_S(\mu_1)} \bigg(1 - \frac{1}{r} - \ln r\bigg) + \bigg( \frac{\Gamma_1}{\Gamma_0} - \frac{\beta_1}{\beta_0} \bigg)(1 - r \ln r) + \frac{1}{2} \frac{\beta_1}{\beta_0} \ln^2 r \bigg], \nn \\
  \eta^{\rm NLL}_{\Gamma_{\rm cusp}}(\mu_1, \mu_2) &= - \frac{1}{2}\frac{\Gamma_0}{\beta_0} \bigg[ \ln r + \frac{\alpha_S(\mu_1)}{4 \pi} \bigg( \frac{\Gamma_1}{\Gamma_0} - \frac{\beta_1}{\beta_0} \bigg)(r - 1) \bigg], \nn \\
  K^{\rm NLL}_{\gamma_J} &= - \frac{1}{2} \frac{\gamma_0}{\beta_0} \ln r \,,
\end{align}
with $r = \frac{\alpha_S(\mu_2)}{\alpha_S(\mu_1)}$, the scales $\mu_H = \Tau_1^{\max}$, $\mu_S = \Tau_1^{\cut}$ and $\mu_J = \sqrt{\mu_H \mu_S}$.
The kinematics-dependent terms are given by
\begin{align}
  Q_a = p_a \, Q_{HH} \, e^{Y_{HH}},\qquad Q_b =  p_b \, Q_{HH} \, e^{-Y_{HH}} ,\qquad Q_J = 2\, p_J\, E_J\,,
\end{align}
where $Q_{HH}$ is the invariant mass of the di-Higgs system, $Y_{HH}$ its rapidity, $p_{a,b}$ are the incoming momenta,
$p_J$ is the momentum and $E_J$ the energy of the jet in the final state (in the frame in which the Higgs pair system
has $Y_{HH} = 0$).
The cusp and noncusp anomalous dimensions are given by
\begin{align}\label{eq:cuspnoncuspad}
  &\Gamma_0 = 4\,, \qquad \Gamma_1 =
  4 \bigg[ \bigg( \frac{67}{9} - \frac{\pi^2}{3} \bigg)C_A - \frac{20}{9}T_F n_f \bigg]\,, \nn \\
  &\gamma_0 = 12 C_F + 2 \beta_0\,, \qquad \beta_0 = \frac{11}{3}C_A - \frac{4}{3}T_F n_f\,, \nn \\
  &\beta_1 = \frac{34}{3}C^2_A - \frac{10}{3}C_A n_f -2 C_F n_f\,.
\end{align}
With $U_1'$ we denote the first derivative of $U_1(\Phi_1,\Tau_1)$ with respect
to $\Tau_1$, and with $U'^{(1)}_1$ and $U^{(1)}_1$ their $\ord{\alpha_S}$
expansions, respectively.

All the non-projectable regions of $\Phi_1$ and $\Phi_2$, due to events that
e.g.\ result in an invalid flavour projection or come from points in the
phase space not covered by the $\Tau_0$-preserving mapping, are included in
the event samples with $1$ or $2$ additional emissions below the resolution
cutoffs. We assign them the following cross sections,
\begin{align}
  \frac{\dsigMC_{1}}{\df\Phi_{1}} (\Tau_0 \le \TauZcutns; \Tau_{1}^\cut) &= (B_1+V_1)(\Phi_1)\, \overline{\Theta}^{\mathrm{FKS}}_{\mathrm{map}}(\Phi_1) \, \theta(\Tau_0<\TauZcutns)\,, \label{eq:1belowtau0}
\end{align}
\begin{align}
  \frac{\dsigMC_{\geq 2}}{\df\Phi_{2}} (\Tau_0 > \TauZcutns; \Tau_{1} \le \Tau_{1}^\cut) &= B_2(\Phi_2)\, \overline{\Theta}_{\mathrm{map}}^\Tau(\Phi_2) \, \theta(\Tau_1 < \Tau_1^\cut)\,
                                                                                           \theta\left(\Tau_0(\Phi_2) > \TauZcutns\right)\,.  \label{eq:2belowtau1}
\end{align}
The quantity $\Theta^{\mathrm{X}}_{\mathrm{map}}$ encodes the constraints due to the projections in the two mappings:
the FKS map in the case of the $\Phi_1 \to {\Phi}_0$ projection and  the
$\Tau_0$-preserving map for the $\Phi_2 \to {\Phi}_1$ projection.
The overlined versions represent their complements.

%%%%%%%%%%%%%%%%%%%%%%%%%%%%%%%%%%%%%%%%%%%%%%%%%%%%%%%%%%%%%%%%%%%%%%%%%%%%%%%%
\subsection{$\Tau_0$ resummation}
\label{sec:tau0}
%%%%%%%%%%%%%%%%%%%%%%%%%%%%%%%%%%%%%%%%%%%%%%%%%%%%%%%%%%%%%%%%%%%%%%%%%%%%%%%%
In order to compute our results in the region below $\Tau_0^\cut$, as well as to perform
the $\Tau_0$ resummation, we rely on the factorisation theorem for the
zero-jettiness in the SCET formalism.
For the particular case at hand, it means that we can write the factorised differential
cross section as~\cite{Stewart:2009yx}
\begin{equation}\label{eq:convolutionmess}
  \frac{\de \sigma^{\rm SCET}}{\de \Phi_0 \, \de \Tau_0} =
  H_{{\scriptscriptstyle gg \to HH }}(Q^2,\mu) \int B_g(t_a,x_a,\mu) \,
  B_g(t_b,x_b,\mu) \, S_{gg}\left( \Tau_0 - \frac{t_a+t_b}{Q}, \mu \right) \,
  \de t_a \, \de t_b.
\end{equation}
In the above formula $H_{{\scriptscriptstyle gg \to HH }}$ is the hard function, $B_g$ the beam
function and $S_{gg}$ the soft function.
The hard function is process dependent and contains the corresponding
Born and virtual matrix elements.
The soft function depends on the external partons in the process:
for colour singlet production in the gluon fusion channel, its perturbative component  is derived from that
calculated for the quark channels through Casimir rescaling.
Similarly, the beam functions also depend on whether the process is quark- or
gluon-initiated. They depend on the transverse virtualities $t_{a,b}$, and for $t_{a,b} \gg \Lambda_{\textrm{QCD}}$
 they satisfy an operator product expansion (OPE) in terms of
perturbative collinear matching coefficients and standard parton distribution functions.

Each of these three functions admits a perturbative expansion in powers of the strong coupling and manifests a logarithmic dependence on a single characteristic scale.
The canonical choice of scales that minimises these logarithmic terms is
\begin{equation}\label{eq:canonicalscales}
  \mu_H = Q, \qquad \mu_B = \sqrt{Q\Tau_0}, \qquad \mu_S = \Tau_0.
\end{equation}
Provided this choice of scales is made, there are no leftover large logarithms
in the perturbative expansion of $H$, $B$ and $S$.
However, the factorisation formula requires all the components to be evaluated
at a single common scale  $\mu$. This is achieved by acting with the
renormalisation group evolution (RGE) operator on each function,
which results in the following schematic representation of the resummed $\Tau_0$
spectrum,
% \begin{align}
%   \frac{\de \sigma^{\rm NNLL^\prime}}{\de \Phi_0 \de \Tau_0} =& H_{\scriptstyle{gg\to HH}}(Q^2,\mu_H)\, U_H(\mu_H, \mu) \{\left[ B_g(t_a,x_a,\mu_B)  \otimes U_B(\mu_B, \mu) \right] \nn \\
%                                                         & \times \left[ B_g(t_b,x_b,\mu_B)  \otimes U_B(\mu_B, \mu) \right] \}\otimes \left[S_{gg}(\mu_S)\otimes U_S(\mu_S, \mu)\right],
%   \label{eq:resum}
% \end{align}
\begin{align}
  \frac{\de \sigma^{\rm NNLL^\prime}}{\de \Phi_0 \de \Tau_0} =&
  H_{{\scriptscriptstyle gg \to HH }}(Q^2,\mu_H)\, U_H(\mu_H, \mu)
                                                           \nonumber \\
 & \times \int \de t_a \,\de t_b  \left[ B_g(t_a,x_a,\mu_B) \otimes U_B(\mu_B, \mu) \right]
    \left[B_g(t_b,x_b,\mu_B) \otimes U_B(\mu_B,\mu) \right]\nonumber \\
 & \qquad \times \left[S_{gg}( \Tau_0 - \frac{t_a+t_b}{Q},\mu_S)\otimes U_S(\mu_S, \mu)\right],
  \label{eq:resum}
\end{align}
where we abbreviated the convolution over internal variables using the symbol $\otimes$.
In the formula above the large logarithms arising from  ratios of disparate
scales in Eq.~\eqref{eq:convolutionmess} have been resummed by the RGE factors $U_i(\mu_i,\mu)$.

At NNLL$^\prime$ accuracy the anomalous dimensions appearing in the evolution
factors need to be known at 2- and 3-loop order for the
noncusp~\cite{Berger:2010xi} and cusp terms~\cite{Moch:2004pa, Vogt:2004mw,
  Korchemsky:1987wg},
respectively. Similarly, the QCD beta function~\cite{Tarasov:1980au,
  Larin:1993tp} is required to be known at 3-loop
order.
In addition, the hard, beam and soft functions have to be computed at 2-loop order.
The necessary soft function has been computed at 2-loops in
Refs.~\cite{Kelley:2011ng,Monni:2011gb}.
The hard function is taken from Refs.~\cite{deFlorian:2013uza}
and~\cite{Grigo:2014jma}, translating the result from the Catani scheme to the
$\overline{\rm MS}$ scheme, which we use in \geneva.
The details of this calculation can be found in Appendix \ref{appendix_A}.
Finally, the beam functions are known up to
3-loops~\cite{Gaunt:2014cfa, Ebert:2020unb}.

In order to extend the resummation accuracy of \geneva
to N$^3$LL, the noncusp and cusp anomalous dimensions~\cite{Berger:2010xi,vanRitbergen:1997va,vonManteuffel:2020vjv}, the QCD
beta function~\cite{vanRitbergen:1997va} and the running of the strong coupling
have to be included at one order higher.

\subsubsection{Choice of scales and their impact on observables}
\label{sec:profileScales}
\begin{figure}[tp]
  \begin{center}
    \includegraphics[width=\rescaletwoplots]{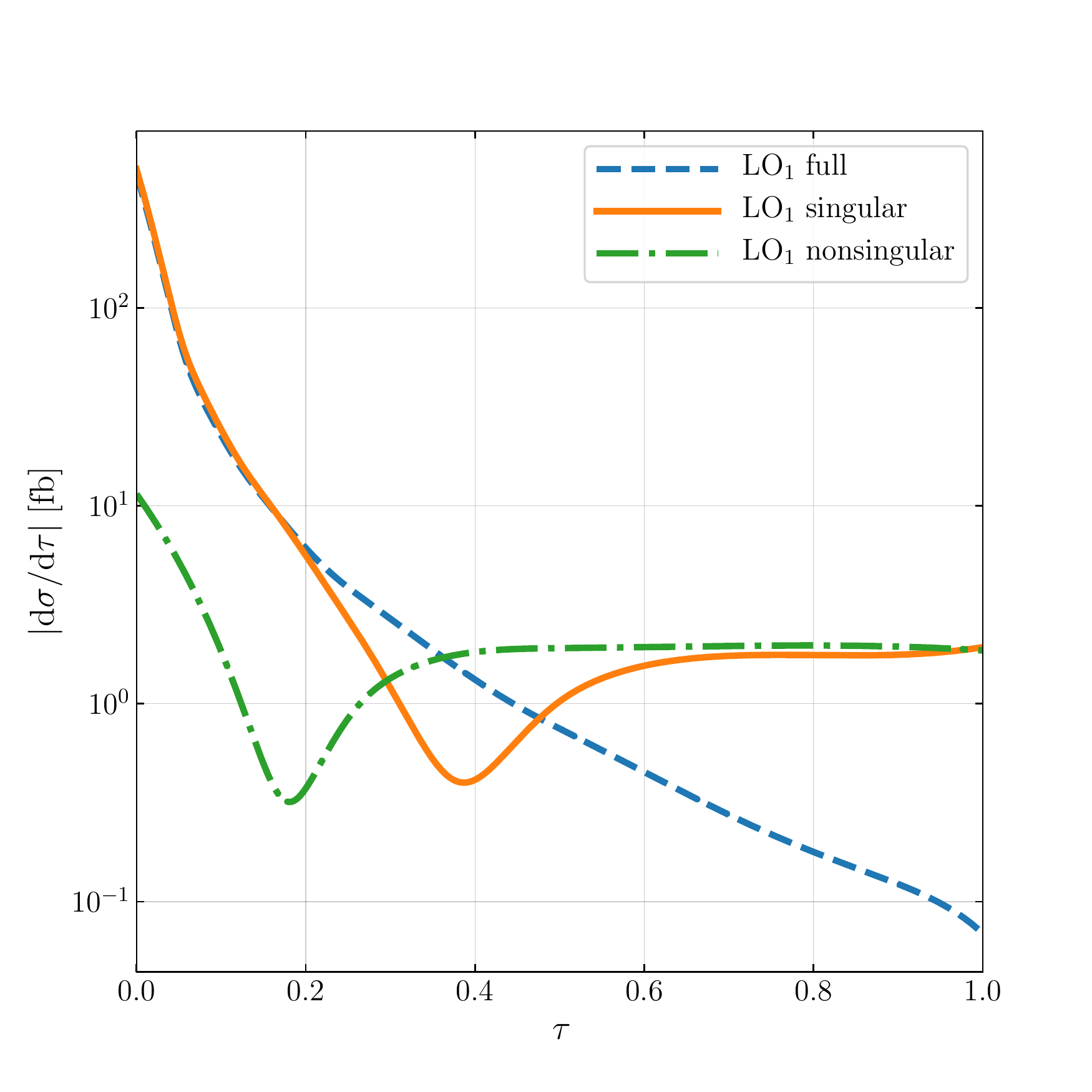}
    \includegraphics[width=\rescaletwoplots]{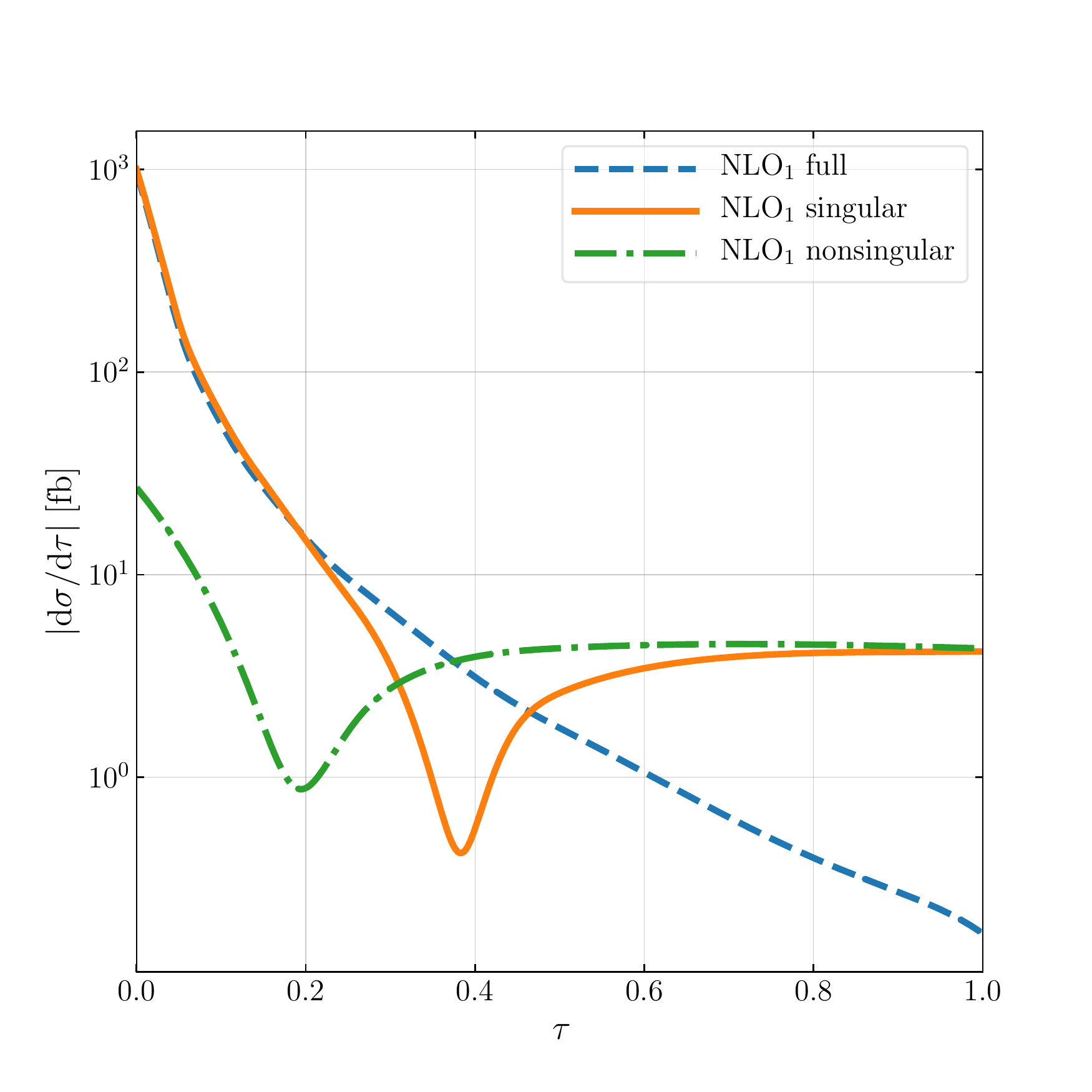}
  \end{center}
  \spaceabovefigurecaption
  \caption{\label{fig:tau0_fo_sing_nonsing}Comparison between the absolute values of the fixed-order
    distribution, of the expansion of the resummed
    contribution up to $\ord{\alpha^2_S}$ (Singular) and of their difference
    (Nonsingular), as a function of $\tau$ at LO$_1$ (left) and NLO$_1$ (right).
  }
  \spacebelowfigurecaption
\end{figure}

In this section we describe the procedure we use to set all the scales entering
both the FO and the resummed parts of our calculation.
To be able to smoothly match the regions where the logarithms of $\tau =
\Tau_0/Q$ are large and where they are sub-dominant, we need to turn off the
resummation around a value of $\mathcal{T}_0\sim Q$, for some hard scale $Q$.
We choose $Q$ to be equal to the FO scale, which in turn equals the
invariant mass of the Higgs boson pair, $M_{HH}$.
For larger values of $\Tau_0$, since the cross section is well approximated by
the FO result, it is important to switch off the resummation as $\Tau_0
\to Q$. Failing to do so
would spoil the cancellation between the
singular and the nonsingular terms.
In the SCET approach, where the resummation is carried out via RGE,
this can be achieved by evolving the soft and beam
functions to the same common nonsingular scale $\mu_{\rm NS} = M_{HH} $ as that of the  hard
function, which is always computed at $\mu_{\rm NS}$.  The evolution of the soft
and beam functions is done by using profile scales $\mu_S(\Tau_0)$ and
$\mu_B(\Tau_0)$. These conventions have been introduced in
Ref.~\cite{Berger:2010xi}, and are given by
\begin{align}
  \mu_S(\Tau_0) &= \mu_{\rm NS} \, f_{\rm run}(\Tau_0/Q), \nn \\
  \mu_B(\Tau_0) &= \mu_{\rm NS} \, \sqrt{f_{\rm run}(\Tau_0/Q)},
\end{align}
where $f_{\rm run}$ is defined as
\begin{align}
  f_{\rm run}(x) = \begin{cases} x_0 [ 1 + ( x/(2x_0))^2 ] & x \leq 2x_0, \\
    x & 2x_0 < x \leq x_1, \\
    x + \frac{(2-x_1-x_2)(x-x_1)^2}{2(x_2-x_1)(x_3-x_1)} & x_1 < x \leq x_2, \\
    1 - \frac{(2-x_1-x_2)(x-x_3)^2}{2(x_3-x_1)(x_3-x_2)} & x_2 < x \leq x_3, \\
    1 & x_3 < x. \end{cases}
\end{align}
This functional form ensures the canonical scaling, given by Eq.~\eqref{eq:canonicalscales},
between $x_0$ and $x_1$ and switches off the resummation above $x_3$.
The region below $2\,x_0$ corresponds to the region where we freeze the running
of all couplings to avoid the Landau pole.
The point $x_2$ corresponds to an inflection point
in the profile function $f_{\rm run}$.
In Fig. \ref{fig:tau0_fo_sing_nonsing} we compare the absolute sizes of the singular and nonsingular
contributions to the cross section as functions of $\tau$ at $\rm LO_1$ and
$\rm NLO_1$ accuracy, where LO$_1$ and NLO$_1$ refer to the order relative to
the partonic phase space with one extra emission. By default we set the profile
parameters to
\begin{equation}
  x_0 = \frac{1 \, \GeV}{Q}, \qquad \{x_1,x_2,x_3 \} = \{0.2, 0.275, 0.35 \}.
\end{equation}
The values of $x_1$ and $x_3$ are chosen at the points where
FO and singular contributions are of similar size and where the
nonsingular contribution becomes dominant, respectively.

We obtain theoretical uncertainties for the FO prediction by varying the central
scale $\mu_{\rm NS}$ up and down by a factor of two and taking the maximal
absolute deviation from the central value as a measure of uncertainty.
For the resummed case we vary the central choices for the profile scales $\mu_S$
and $\mu_B$ independently, as e.g.\ detailed in~\cite{Alioli:2019qzz}, keeping
$\mu_H = \mu_{\rm NS}$ fixed.
We include also two more profiles where all the $x_i$ are varied by $\pm 0.05$
simultaneously, while keeping all the other scales at their central values.
In total we get six profile variations and take the maximal absolute deviation
in the result from the central value as the resummation uncertainty.
The total uncertainty is then given by the quadrature sum of the resummation and FO uncertainties.

Due to the dependence on $\Tau_0$ of the profile scale $\mu$, the integral over
the spectrum is not equal to the
cumulant,
\begin{equation}
  \int_{0}^{\Tau^{\rm max}_0} \frac{\de \sigma^{\rm NNLL^\prime}}{\de \Phi_0 \de \Tau_0}(\mu(\Tau_0)) \; \de \Tau_0 = \frac{\de \sigma^{\rm NNLL^\prime}}{\de \Phi_0}(\Tau^{\rm max}_0, \mu(\Tau^{\rm max}_0)) + \ord{\rm N^3LL},
\end{equation}
where $\Tau^{\rm max}_0$ is the upper kinematical limit.
While the difference is of higher order it can be numerically large~\cite{Bertolini:2017eui}.
When matching this resummed result to a
FO calculation, this can cause a sizeable difference between the total
matched cumulant and a purely fixed-order NNLO cross section, even in the
absence of subleading power corrections.
To obviate this problem, we add an additional higher-order term to our spectrum,
\begin{align}
  \label{eq:improvedXS}
  \frac{\de \sigma^{\rm improvedXS}}{\de \Phi_0 \de \Tau_0}(\mu(\Tau_0)) &= \frac{\de \sigma^{\rm NNLL^\prime}}{\de \Phi_0 \de \Tau_0}(\mu(\Tau_0)) \nn \\
                                                                         &+ p \, \mathcal{K}(\Tau_0, \Phi_0)\bigg[ \frac{\de}{\de \Tau_0}\frac{\de \sigma^{\rm NNLL^\prime}}{\de \Phi_0}(\Tau_0,\mu_h(\Tau_0)) - \frac{\de \sigma^{\rm NNLL^\prime}}{\de \Phi_0 \de \Tau_0}(\mu_h(\Tau_0)) \bigg],
\end{align}
where $\mu_h(\Tau_0)$ is a dedicated profile scale and $\mathcal{K}(\Tau_0, \Phi_0)$ is a smooth function defined as
\begin{align}
  \label{eq:kfactor}
  \mathcal{K}(\Tau_0, \Phi_0) = \frac{1}{2} - \frac{1}{2} \tanh \bigg[ 32\, \bigg( \frac{\Tau_0}{M_{HH}} - \frac{1}{4} \bigg) \bigg].
\end{align}
Note that by construction the additional term in Eq. \eqref{eq:improvedXS} is of
higher order, consequently the $\rm NNLL^\prime$ accuracy of the spectrum is not
spoiled.
Moreover, its effects are limited to the resummation region, since
$\mu_h(\Tau_0) = Q$ in the FO region so that the difference in the
square brackets of Eq.~\eqref{eq:improvedXS} vanishes.
The function $\mathcal{K}(\Tau_0, \Phi_0)$ is chosen such
that it tends to zero for large values of $\Tau_0$, and that the
effects of the induced higher-order terms are compatible with the scale
uncertainties of the original spectrum in the peak region.
Consequently, the additional
higher-order terms induced by this procedure contribute mostly in the
peak and transition regions, where they are expected to be larger.
Lastly, we tune the $p$ values to ensure that we recover the total inclusive
cross section upon integration.
In order to do so,
we fix the value of $p$ by requiring that the integral over this modified
version of the spectrum is equal to that of the cumulant,
\begin{align}
  \label{eq:pvalue}
  p = \frac{ \int \de \Phi_0 \int \de \Tau_0 \bigg[ \frac{\de}{\de \Tau_0}\frac{\de \sigma^{\rm NNLL^\prime}}{\de \Phi_0}(\Tau_0,\mu_h(\Tau_0)) - \frac{\de \sigma^{\rm NNLL^\prime}}{\de \Phi_0 \de \Tau_0}(\mu_h(\Tau_0)) \bigg]}{ \int \de \Phi_0 \int \de \Tau_0 \bigg[  \frac{\de}{\de \Tau_0}\frac{\de \sigma^{\rm NNLL^\prime}}{\de \Phi_0}(\Tau_0,\mu_h(\Tau_0)) - \frac{\de \sigma^{\rm NNLL^\prime}}{\de \Phi_0 \de \Tau_0}(\mu_h(\Tau_0)) \bigg]\,\mathcal{K}(\Tau_0, \Phi_0) }.
\end{align}
One must however pay attention to the fact that both the  value  of $p$ in
Eq.~\eqref{eq:pvalue} and the $\mathcal{K}(\Tau_0, \Phi_0)$ factor in
Eq.~\eqref{eq:kfactor} are obtained integrating over the Born variables
$\Phi_0$. In particular, there is a nontrivial interplay between the Higgs pair
invariant mass $M_{HH}$ and the definition of $\mathcal{K}(\Tau_0,
\Phi_0)$. Since the $M_{HH}$ distribution presents a maximum around $4 m_H \sim
450-500 \; \GeV$, but spans over a wide range, using a single value of $p$
across all the possible values of $M_{HH}$ has a sizable effect  on
the predicted $M_{HH}$ differential distribution, despite the fact that the correct inclusive cross
section is obtained by construction.
We discuss this issue in more detail in the validation against
the NNLO result, in Sec.~\ref{sec:validation}.

\subsubsection{Partonic predictions}
\label{sec:NNLO}
\begin{figure}[tp]
  \begin{center}
    \includegraphics[width=\rescalethreeplots]{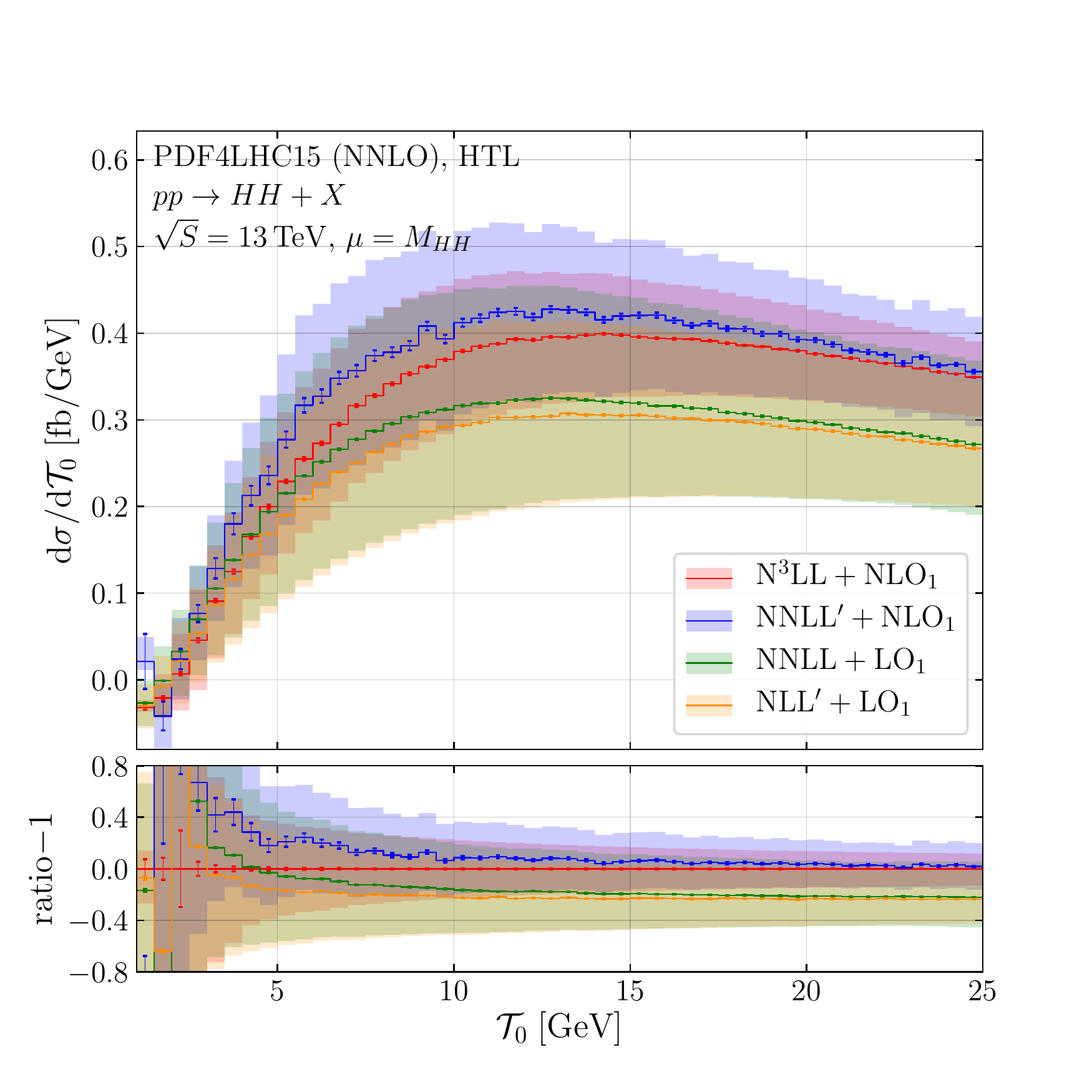}
    \includegraphics[width=\rescalethreeplots]{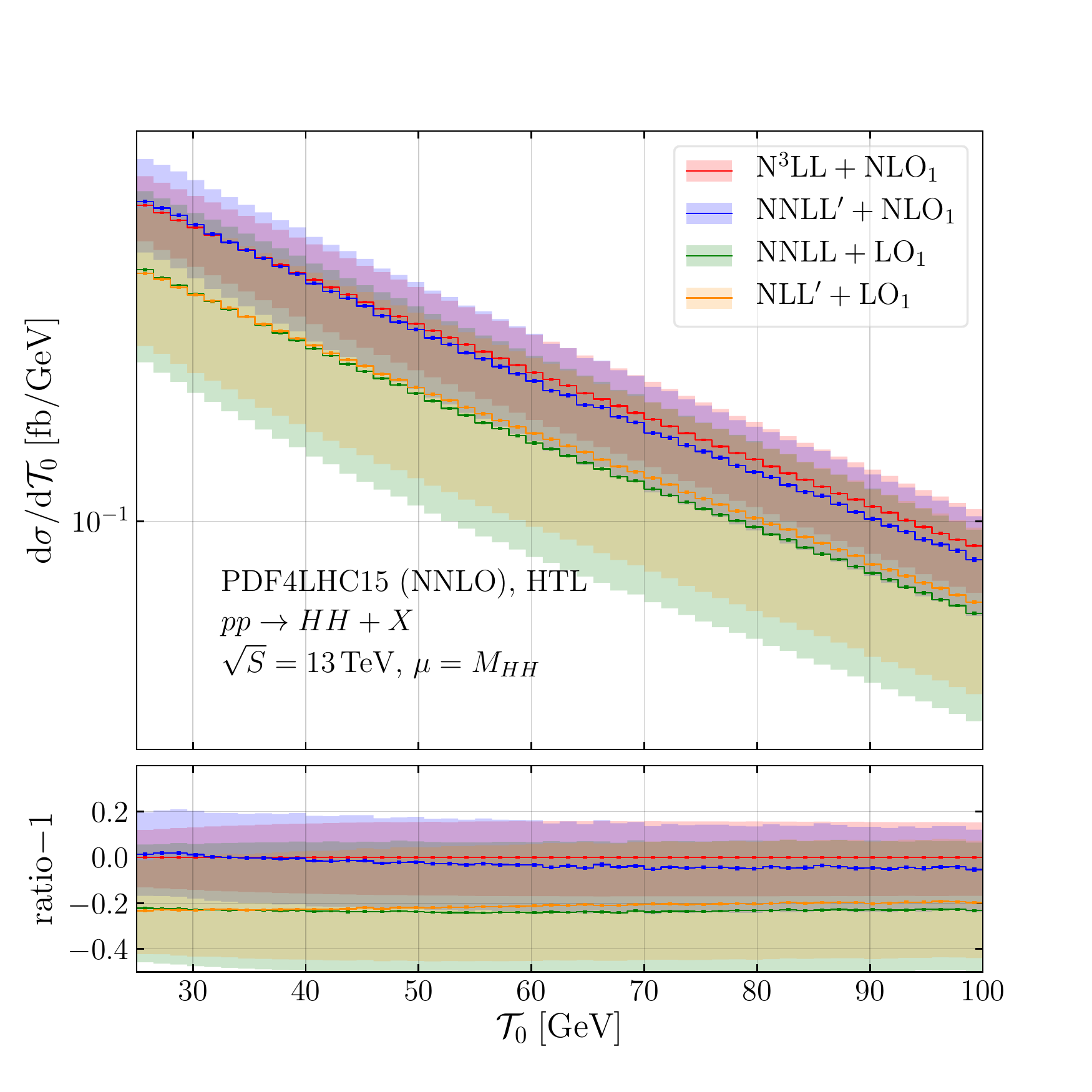}
    \includegraphics[width=\rescalethreeplots]{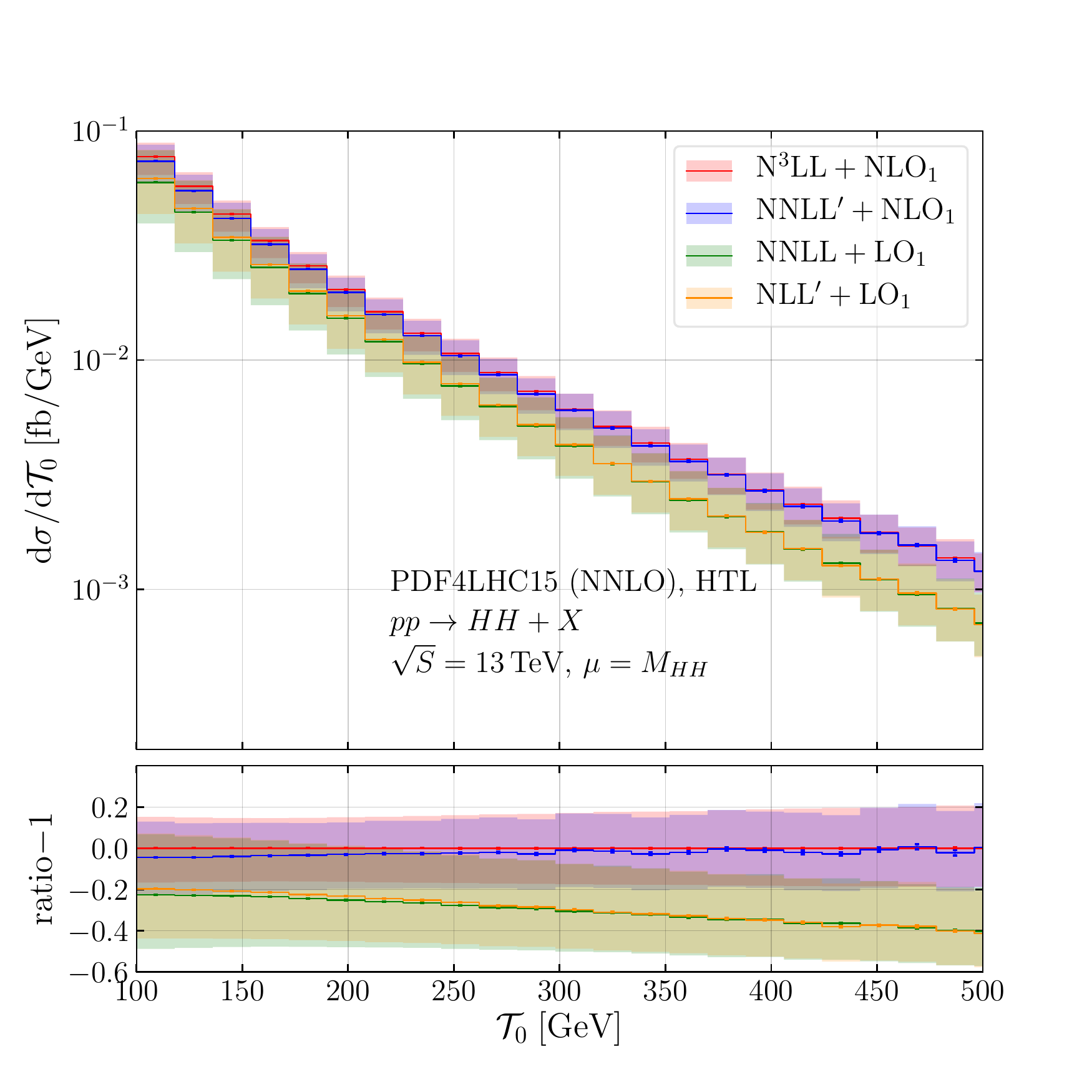}
  \end{center}
  \spaceabovefigurecaption
  \caption{\label{fig:convergence_tau0}Resummed predictions matched to the appropriate fixed-order results at different accuracies for the $\Tau_0$ distribution in the peak (left), transition (centre) and tail (right) region.
  }
  \spacebelowfigurecaption
\end{figure}

We now discuss the numerical impact of the $\Tau_0$ resummation as well as that
of the choice of the resummation cut $\TauZcutre$.
In Fig.~\ref{fig:convergence_tau0} we show resummed predictions for the $\Tau_0$ distribution divided into
three different regions: peak, transition and tail.
We present the resummed results at different resummation orders matched to the
appropriate FO calculations: NLL$^\prime$+LO$_1$,
NNLL+LO$_1$, NNLL$^\prime$+NLO$_1$ and N$^3$LL + NLO$_1$.
In this, and in all the following plots, we report both the statistical errors
due to the Monte Carlo integration, which appear as vertical bars, as well as
the scale variation band, obtained by the procedure previously discussed. The
lower insets of the plots show instead the normalised relative ratios between
the curves.
As expected, the peak region at small $\Tau_0$ is where the
resummation has the largest impact, which is reflected by a large spread among
the predictions at different resummation accuracies.
In the transition and tail regions  the
difference between the various predictions is driven by the FO accuracy, with LO$_1$ results being consistently
smaller than the NLO$_1$ ones across the whole range.
We observe a reasonable convergence of the perturbative predictions:
the scale variations  at $\rm N^3LL + NLO_1$
and $\rm NNLL^\prime + NLO_1$ are smaller than those at lower
orders, especially in the peak and transition regions.
We notice however that, contrary to what one might have anticipated, resummation
effects are still visible up to values of $\Tau_0\lesssim 300$~GeV in the small
difference between the N$^3$LL and the NNLL$^\prime$ results, at the order of
a few percent.
This can be explained by the fact that the actual resummed variable is $\tau_0 =
\mathcal{T}_0/M_{HH}$, which, in this particular
process, can be small even for relatively large values of $\mathcal{T}_0$, when
$M_{HH}$ becomes very large.

\begin{figure}[tp]
  \begin{center}
    \includegraphics[width=\rescalethreeplots]{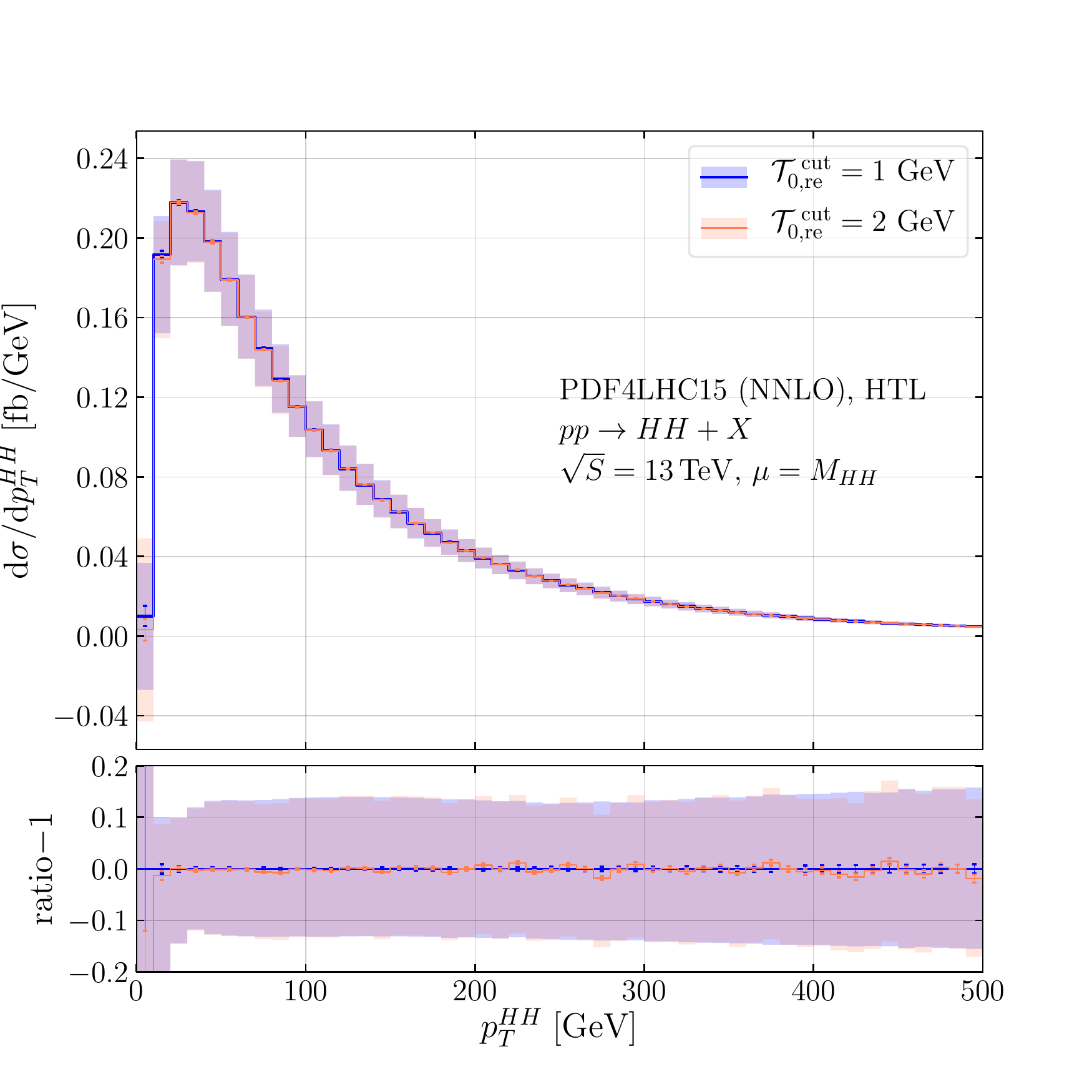}
    \includegraphics[width=\rescalethreeplots]{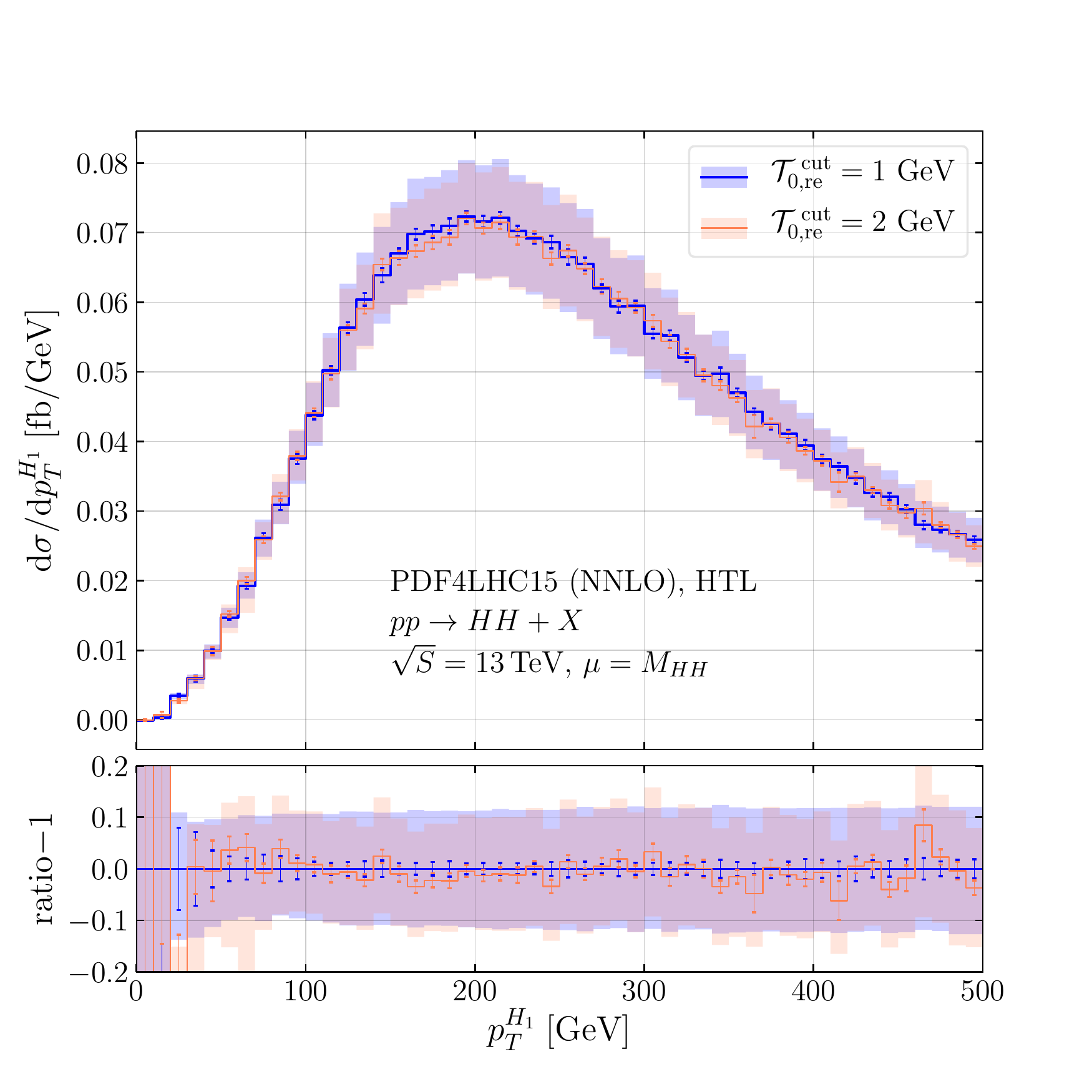}
    \includegraphics[width=\rescalethreeplots]{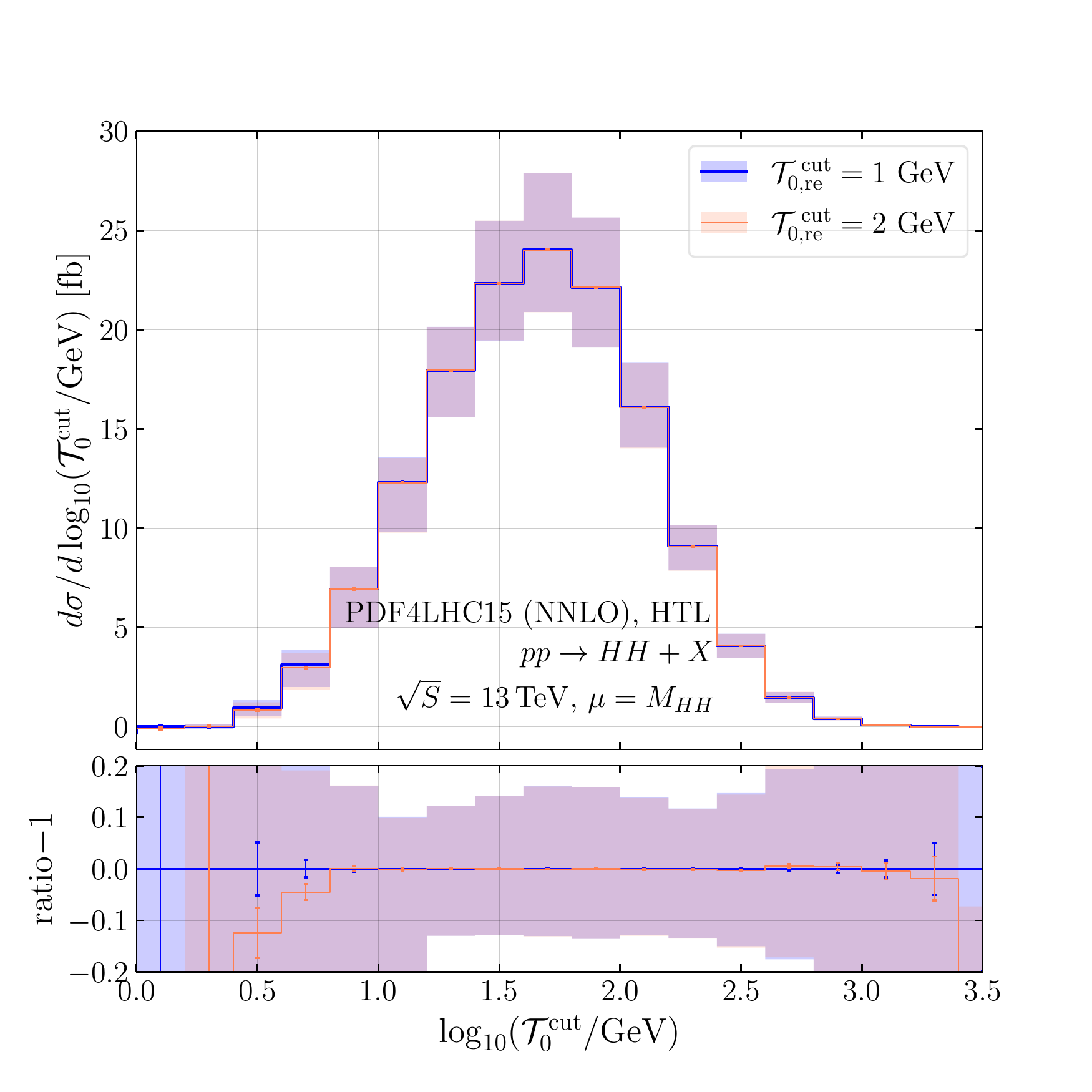}
  \end{center}
  \spaceabovefigurecaption
  \caption{\label{fig:independece_tau0cutres}Comparison between $\TauZcutre = 1,\: 2 \; \GeV$ at fixed
    $\TauZcutns = 0.5 \; \GeV$ for $p_T^{HH}$ (left), $p_T^{H_1}$(centre) and
    $\Tau_0$ (right) distributions.
  }
  \spacebelowfigurecaption
\end{figure}

As discussed in the previous section, we use different cuts
for the \zero-jet resolution for the resummed and the nonsingular
components.
We remark that a variation in the value of $\TauZcutre$ only amounts to
shifting part of the resummed contribution from the \mbox{\zero-jet} bin to the
spectrum, and vice versa.
In general, the resummed calculation might be problematic when the soft
and beam scales reach small values of the order $\lqcd$, due to the running of the strong
coupling. The introduction of profile scales that smoothly turn off the
divergence, freezing the soft scale and preventing it from approaching $\lqcd$,
partially solves this problem, but renders the perturbative resummed calculation
unreliable in that extreme region. Moreover, $\TauZcutre$ is eventually tied to the
starting scale of the parton shower inside the \zero-jet bin.  Therefore, for both
of the previous reasons, it is advisable not to push the
$\TauZcutre$ to too small values.
In Fig.~\ref{fig:independece_tau0cutres} we study the dependence of the \geneva
partonic results on the choice of $\TauZcutre$ for a fixed
$\TauZcutns = 0.5$~GeV. As expected, this choice does not impact in a
statistically significant way the distributions shown, which are the
transverse momentum of the Higgs pair $p_T^{HH}$, the transverse momentum of the
hardest Higgs boson $p_T^{H_1}$ and the zero-jettiness, respectively.

%%%%%%%%%%%%%%%%%%%%%%%%%%%%%%%%%%%%%%%%%%%%%%%%%%%%%%%%%%%%%%%%%%%%%%%%%%%%%%%%
\subsection{Nonsingular and power-suppressed corrections}
%%%%%%%%%%%%%%%%%%%%%%%%%%%%%%%%%%%%%%%%%%%%%%%%%%%%%%%%%%%%%%%%%%%%%%%%%%%%%%%%
\begin{figure}[tp]
  \begin{center}
    \includegraphics[width=\rescaletwoplots]{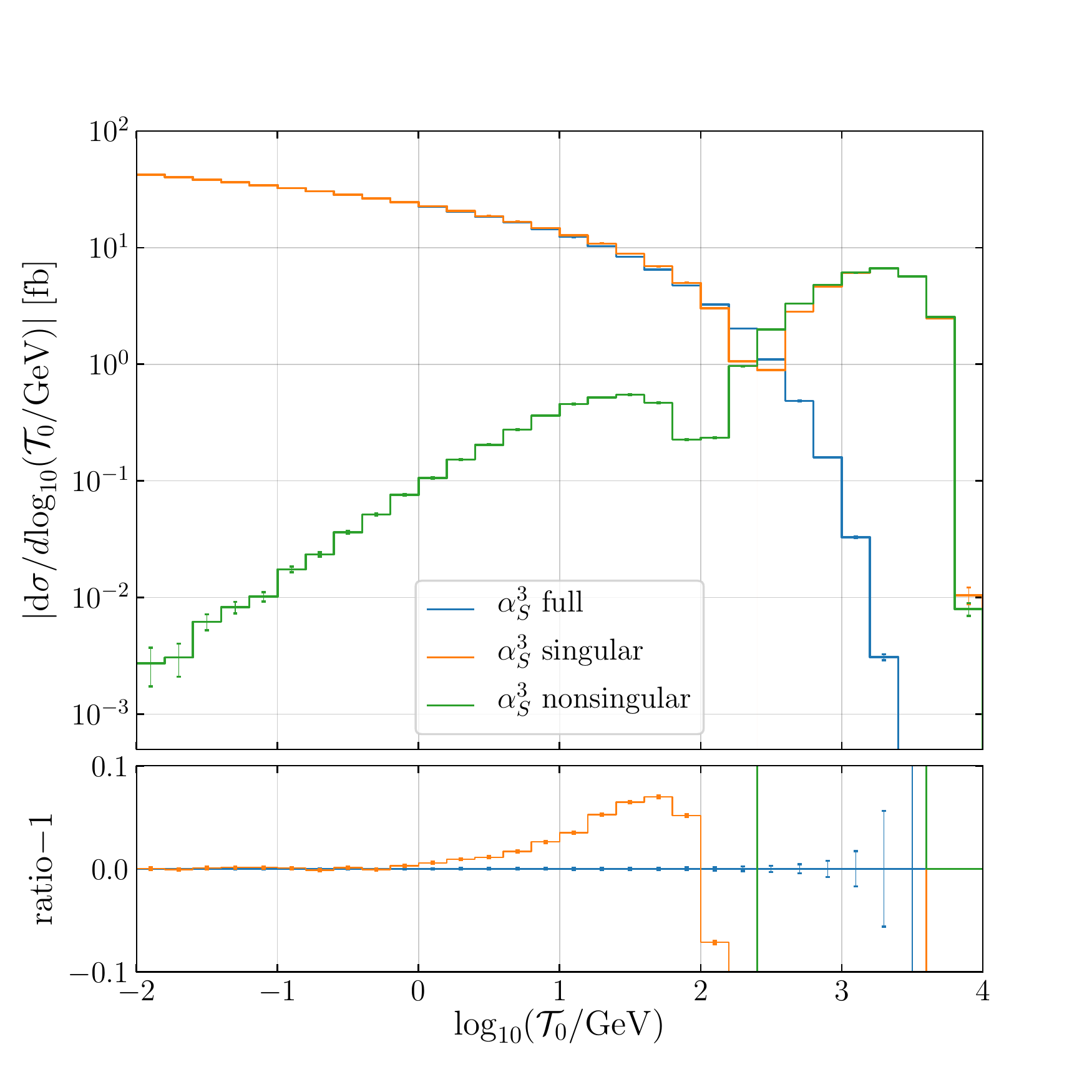}
    \includegraphics[width=\rescaletwoplots]{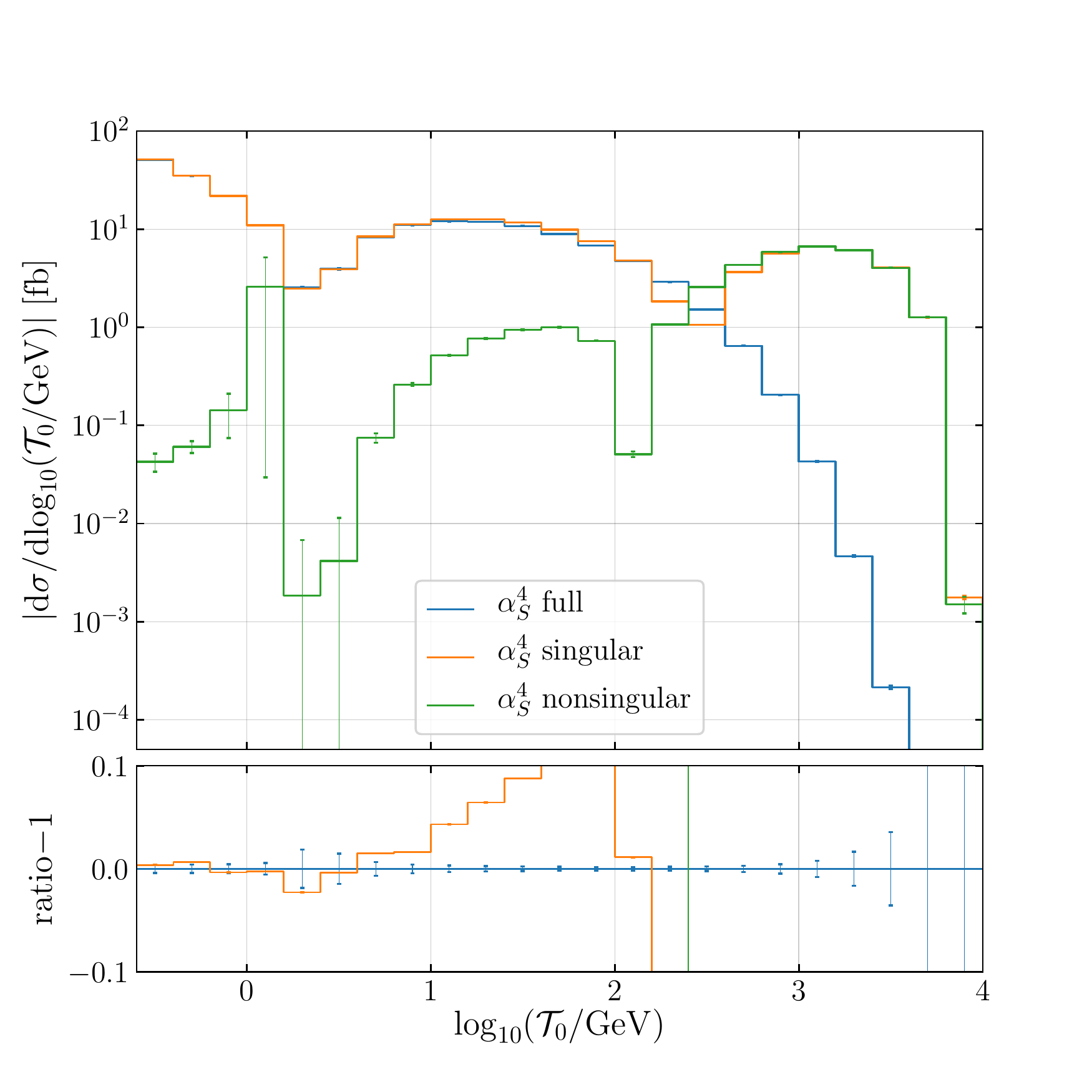}
  \end{center}
  \spaceabovefigurecaption
  \caption{\label{fig:LO1_NLO1_cancellation}Singular and nonsingular contributions to the double Higgs production cross section as a function of $\Tau_0$ at NLO (left) and NNLO (right).
  }
  \spacebelowfigurecaption
\end{figure}

\begin{figure}[htp]
  \begin{center}
    \includegraphics[width=8.0cm]{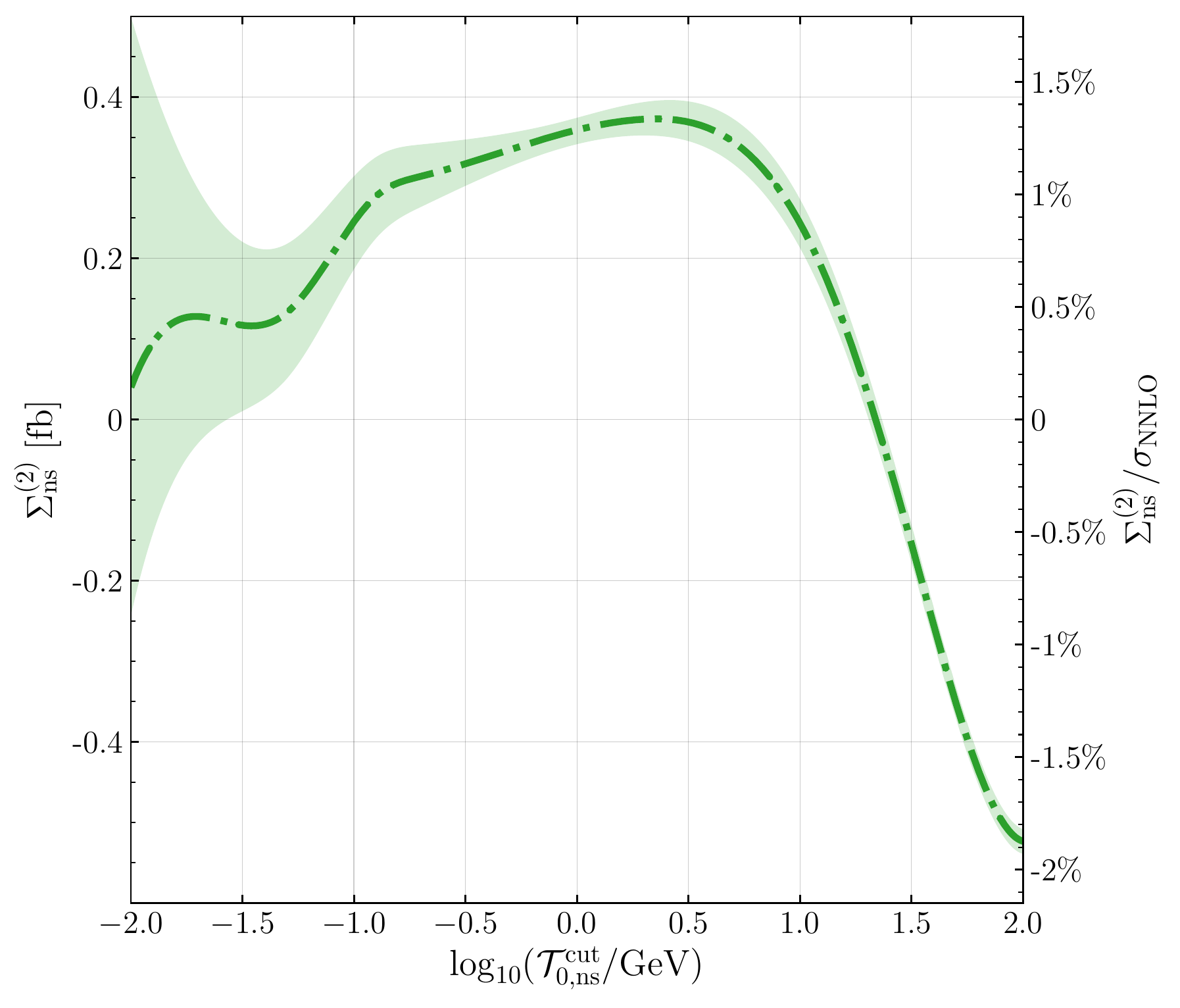}
  \end{center}
  \spaceabovefigurecaption
  \caption{\label{fig:Tau0_nonsing_cumulant}The neglected $\ord{\alpha_S^2}$
    nonsingular contribution to the $\Tau_0$ cumulant $\Sigma_{\NS}^{(2)}$ as a
    function of $\TauZcutns$. The green band represents the statistical uncertainty.
  }
  \spacebelowfigurecaption
\end{figure}

In the \geneva framework, all the contributions below the cut on the 0-jet resolution variable are given in Eq.~\eqref{eq:0full}.
This expression is NNLO accurate and fully differential in the phase space $\Phi_0$.
In principle one could use a local NNLO subtraction for the implementation of such terms.
Alternatively, one can approximate, up to
power corrections, the expression in
Eq.~\eqref{eq:0full} with
\begin{align}
  \label{eq:0full_mod}
  \frac{\widetilde{\de \sigma^{\rm MC}_0}}{\de \Phi_0}(\TauZcutre, \TauZcutns) =& \, \frac{\de \sigma^{\rm NNLL^\prime}}{\de \Phi_0}(\TauZcutre) - \frac{\de \sigma^{\rm NNLL^\prime}}{\de \Phi_0}(\TauZcutns) \bigg|_{\rm NLO_0} + (B_0 + V_0)(\Phi_0) \nn \\
                                                                                & + \int B_1(\Phi_1) \, \theta( \Tau_0(\Phi_1) < \TauZcutns ) \, \frac{\de \Phi_1}{\de \Phi_0}\,.
\end{align}
This formula only requires a NLO subtraction and the expansion of the resummed
contribution at $\ord{\alpha_S}$ relative to the leading order. It is based on
the fact that the
singular and FO contributions cancel
up to power corrections below the resolution cutoff at $\ord{\alpha_S^2}$.
The cancellation between these two terms as a function of the resolution cutoff
$\Tau_0$ is shown in Fig.~\ref{fig:LO1_NLO1_cancellation} by plotting the
absolute values of their central predictions, both at $\rm LO_1$, which is of
absolute order $\alpha_S^3$, on the left
and the pure $\alpha_S^4$ contribution on the right.
We notice that the nonsingular distribution correctly approaches zero while the
separate FO and singular contributions are diverging, both at order
$\alpha_S^3$ and $\alpha_S^4$. This is despite the appearance of numerical
instabilities in the region where the $\alpha_S^4$ nonsingular changes sign,
around $\Tau_0\sim 1.5$~GeV.
In practice, however, for any finite choice of $\TauZcutns$ there are always
remaining nonsingular power corrections below the cutoff, identified by the
difference between Eq.~\eqref{eq:0full} and \eqref{eq:0full_mod}, which reads
\begin{align}
\label{eq:nonsingcumas2}
  \frac{\de \Sigma^{(2)}_{\rm ns}}{\de \Phi_0}(\TauZcutns) =& - \frac{\de \sigma^{\rm NNLL^\prime}}{\de \Phi_0}(\TauZcutns)\bigg|_{\rm NNLO_0} +  \frac{\de \sigma^{\rm NNLL^\prime}}{\de \Phi_0}(\TauZcutns)\bigg|_{\rm NLO_0} + W_0(\Phi_0) \nn \\
                                                            &+ \int V_1(\Phi_1) \, \theta(\Tau_0(\Phi_1) < \TauZcutns) \, \frac{\de \Phi_1}{\de \Phi_0} \nn \\
                                                            &+ \int B_2(\Phi_2) \, \theta(\Tau_0(\Phi_2) < \TauZcutns) \, \frac{\de \Phi_2}{\de \Phi_0}.
\end{align}
This term scales like a power correction in $\TauZcutns/Q$ and is of
$\ord{\alpha_S^2}$ relative to the Born contribution. We show its absolute size as a function of $\TauZcutns$ in
Fig. \ref{fig:Tau0_nonsing_cumulant}, as well as its relative size as a fraction of
the NNLO cross section computed by
\Matrix~\cite{Grazzini:2017mhc,deFlorian:2016uhr} on the right axis.
For the results presented in this work we choose $\TauZcutns = 0.5 \, \GeV$. The
size of the missing corrections associated to that value is around $1.2\%$ of
the total cross section. These missing contributions only affect events below
the cutoff, and we can
recover the exact NNLO cross section by reweighting these events by this difference.
Note that, while we could have chosen smaller values of $\TauZcutns$ to further
minimise the impact of power corrections, lowering this value has shown to cause
instabilities in the matrix elements used for our calculation.

%%%%%%%%%%%%%%%%%%%%%%%%%%%%%%%%%%%%%%%%%%%%%%%%%%%%%%%%%%%%%%%%%%%%%%%%%%%%%%%%
\section{Details of the calculation and validation of NNLO results}
\label{sec:validation}
%%%%%%%%%%%%%%%%%%%%%%%%%%%%%%%%%%%%%%%%%%%%%%%%%%%%%%%%%%%%%%%%%%%%%%%%%%%%%%%%
We consider the process $pp \to HH + X$ and work in the HTL, taking into account
only the gluon fusion production
channel and requiring two on-shell Higgs bosons in the final state.
The centre-of-mass energy considered is $\sqrt{S} = 13$~TeV, and we use the
following input parameters:
\begin{equation}
   m_H = 125.09 \; \GeV, \qquad   v = 246.32 \; \GeV, \qquad m_t = 173.1 \; \GeV.
\end{equation}
We set both the factorisation and renormalisation scales to the invariant mass
$M_{HH}$ of the Higgs pair,
and we use the $\texttt{PDF4LHC15\_nnlo\_100}$ PDF~\cite{Butterworth:2015oua}
set
from \textsc{LHAPDF~6}~\cite{Buckley:2014ana}, including the corresponding value
of $\alpha_S(M_Z)$. By default, we use the three-loop running of $\alpha_S$ for both \Matrix
and \geneva predictions.
To evaluate the beam functions we use the \texttt{beamfunc} module of
\texttt{scetlib}~\cite{Billis:2019vxg,scetlib}.
We set our resolution cutoffs as $\TauZcutre = 1 \; \GeV$, $\TauZcutns = 0.5
\; \GeV$ and $\Tau_1^{\rm cut} = 1 \; \GeV$, which provide a reasonable
compromise in terms of the size of the neglected power-suppressed terms and
the stability of the singular-nonsingular
cancellation.
\begin{figure}[tp]
  \begin{center}
    \includegraphics[width=\rescaletwoplots]{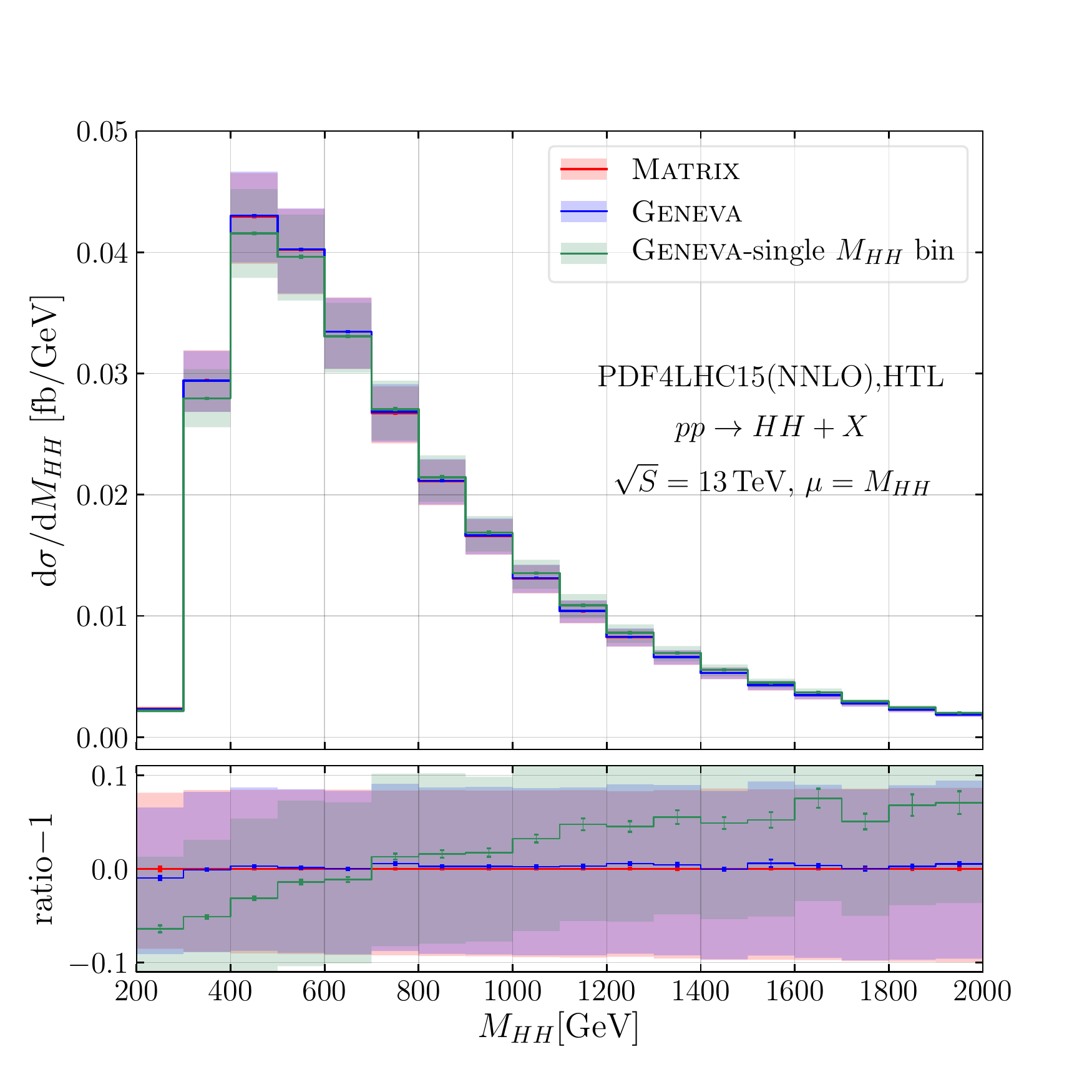}
    \includegraphics[width=\rescaletwoplots]{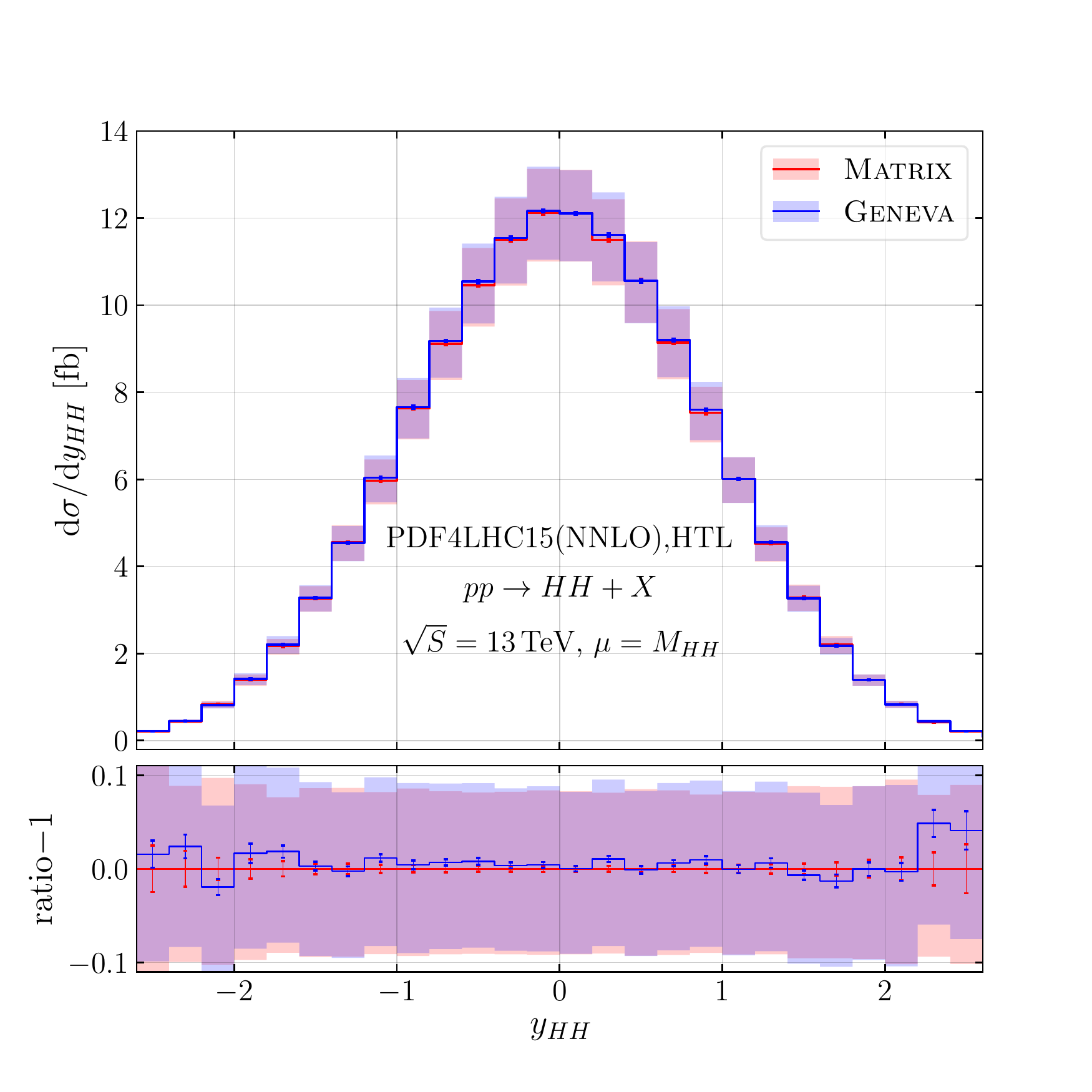}
  \end{center}
  \spaceabovefigurecaption
  \caption{\label{fig:inclusive_dist_higgsPair}Comparison of the distribution
    for the Higgs pair invariant mass (left) and rapidity (right) between \Matrix and \geneva.}
  \spacebelowfigurecaption
\end{figure}
All tree-level matrix elements are calculated using
\textsc{Recola}~\cite{Denner:2017wsf,Denner:2017vms} via a novel custom-built
interface to \textsc{Geneva},
while all the one-loop terms are calculated through the standard
\textsc{Openloops}~\cite{Buccioni:2017yxi} interface already used for
processes previously implemented.

\begin{figure}[t]
  \begin{center}
    \includegraphics[width=\rescaletwoplots]{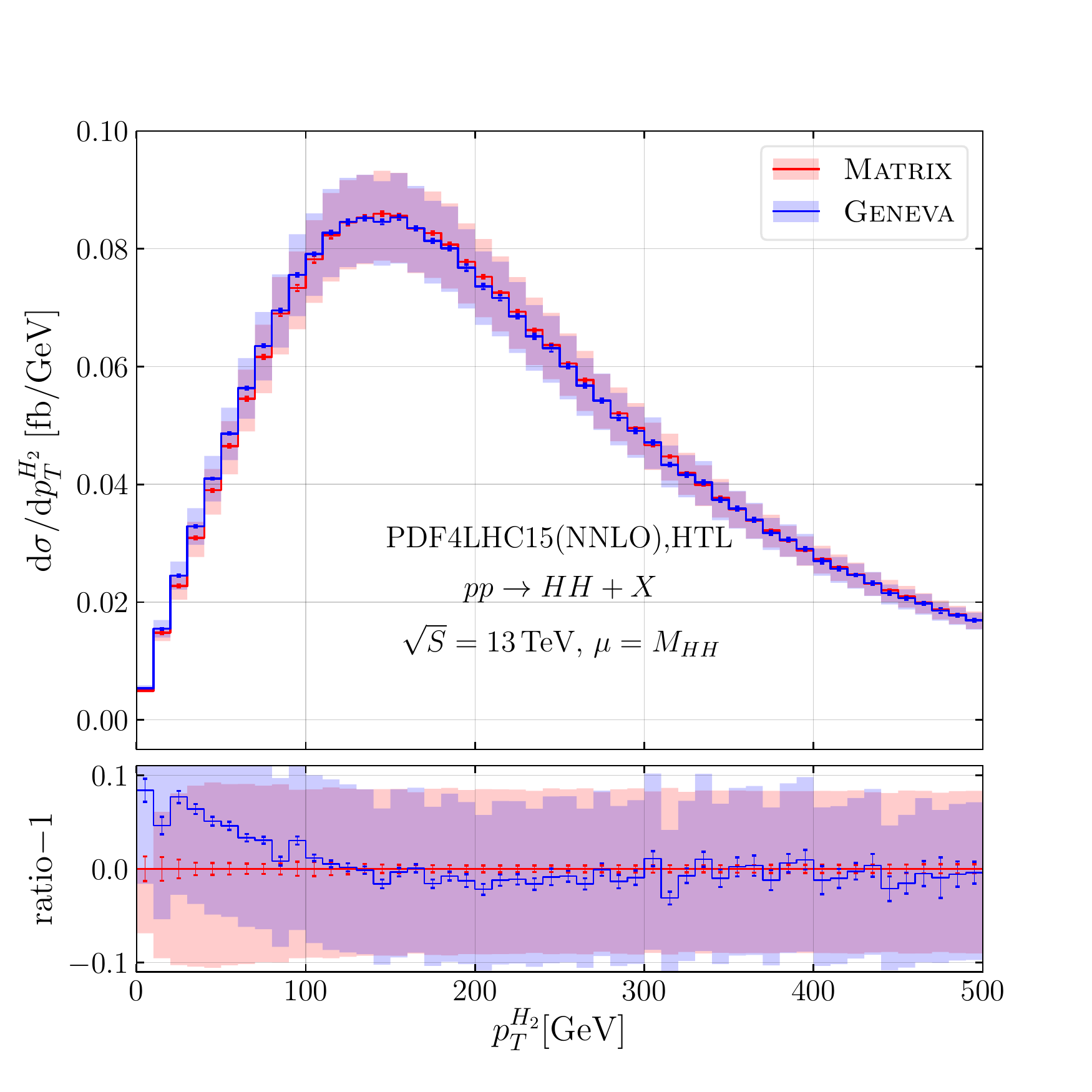}
    \includegraphics[width=\rescaletwoplots]{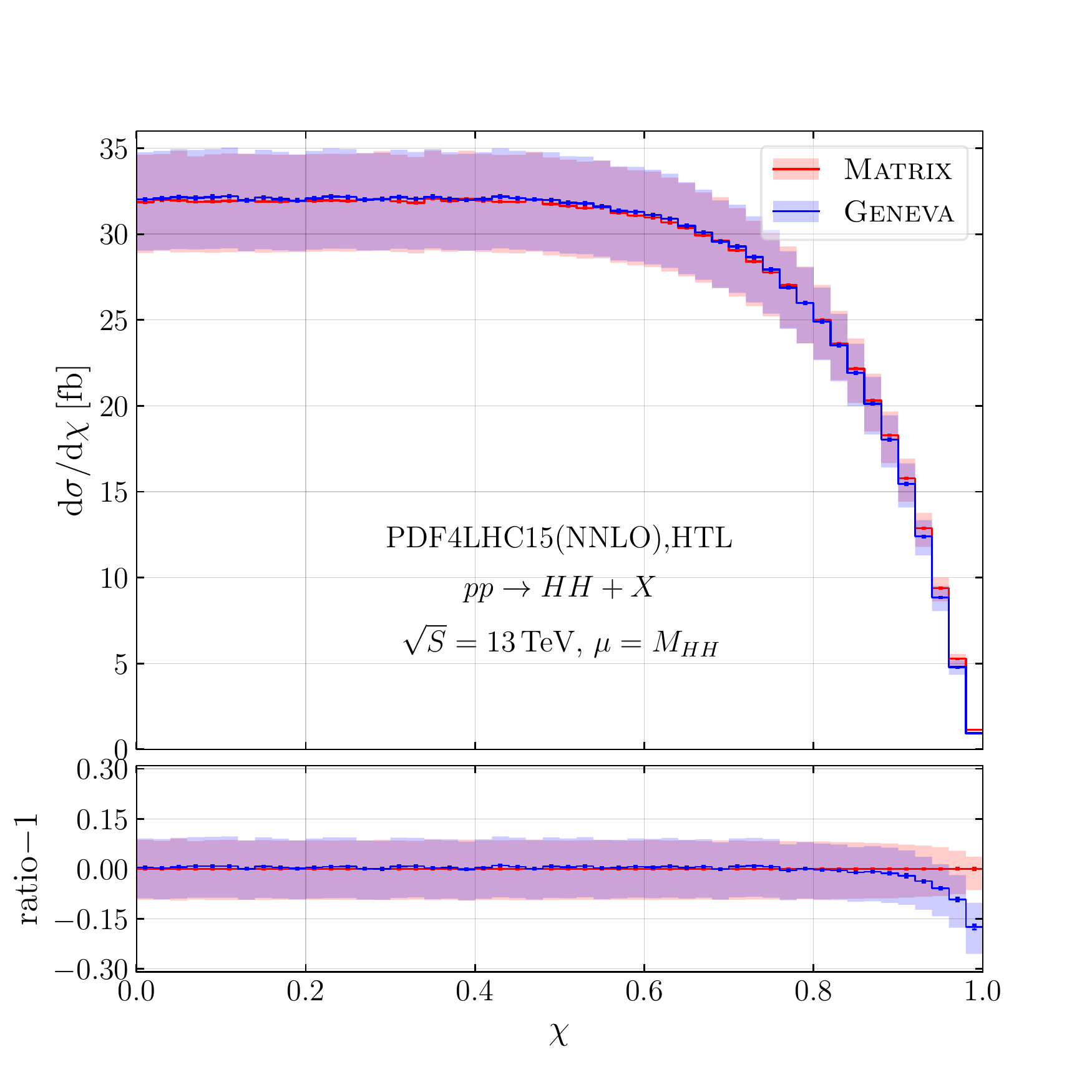}
  \end{center}
  \spaceabovefigurecaption
  \caption{\label{fig:exclusive_dist_higgsPair}Comparison of the distributions
    for the transverse momentum of the softest Higgs boson (left) and the
    absolute value of their scattering angle (right) between \Matrix and
    \geneva.
  }
  \spacebelowfigurecaption
\end{figure}

To validate the NNLO accuracy of the results obtained with \geneva, we compare our results
to an independent calculation implemented in \Matrix~\cite{Grazzini:2017mhc,deFlorian:2016uhr}.
Note that this comparison is done at partonic level, before interfacing to the parton shower.
In Sec.~\ref{sec:shower} we will show the results at the showered level, highlighting how
all the inclusive quantities presented in this section are well preserved by the shower
stage.
We report the result of this comparison in
Fig.~\ref{fig:inclusive_dist_higgsPair} and
Fig.~\ref{fig:exclusive_dist_higgsPair} where we show the invariant mass
$M_{HH}$, the rapidity $y_{HH}$ of the Higgs pair, the transverse momentum of
the softest of the two Higgs bosons and an observable depending on the rapidity difference between the two Higgs bosons,
respectively.
In Fig.~\ref{fig:inclusive_dist_higgsPair},
we show the comparison between \Matrix (red), \geneva (blue) and, in the left
panel, the \geneva line
obtained by using Eq.~(\ref{eq:pvalue}) and Eq.~(\ref{eq:improvedXS}) at face
value (green). As anticipated, higher-order effects do play a
significant role for the invariant-mass distribution of the Higgs pair.
Indeed, we remind the reader that the green and blue lines here differ only by
higher orders. To be precise, the blue line, which corresponds to the default \geneva prediction in
this work, is obtained by computing the value $p$ of Eq.~(\ref{eq:pvalue}) --
and thus Eq.~(\ref{eq:kfactor}) -- in different bins of $M_{HH}$. The reason
for this is given by the fact that, as explained in Sec.~\ref{sec:tau0}, the
resummation we perform is for small $\tau_0$ values, which means that one can
have large resummation effects even at large values of $\Tau_0$ and $M_{HH}$
provided their ratio is small.
We have therefore combined various runs in different $M_{HH}$ bins,
distributed more densely in the peak region  where the cross section is larger.
As can be seen, the difference between these two curves spans a range from
$-5\%$ to roughly $+10\%$. While in general we find a reasonable agreement for
these inclusive distributions within the uncertainty bands of the two
calculations, obtained with the 3-point $\mu_R$ and $\mu_F$ variations,  only
the blue line shows perfect agreement with the \Matrix result.

In Fig. \ref{fig:exclusive_dist_higgsPair} we report the comparison
for the transverse momentum of the softest Higgs boson and the
hyperbolic tangent of the rapidity difference between the two Higgs bosons, defined as
\begin{equation}
  \chi = \tanh \left( \frac{|y_{H_1}-y_{H_2}|}{2} \right),
\end{equation}
where $y_{H_1,H_2}$ are the rapidities of each Higgs boson.
Similarly to what is observed in the previous figure, we notice a good agreement,
except in the regions of small $p_T^{H_2}$ or  large $\chi$.
These differences are expected, as although predictions obtained in the \geneva
framework are NNLO accurate, they include a certain amount of higher-order effects, as
well as power-suppressed corrections, which can lead to these kind of
discrepancies. Note that these power corrections arise from a different physical
treatment of kinematics compared to a FO calculation~\cite{Ebert:2020dfc}.

\begin{figure}[tp]
  \begin{center}
    \includegraphics[width=\textwidth]{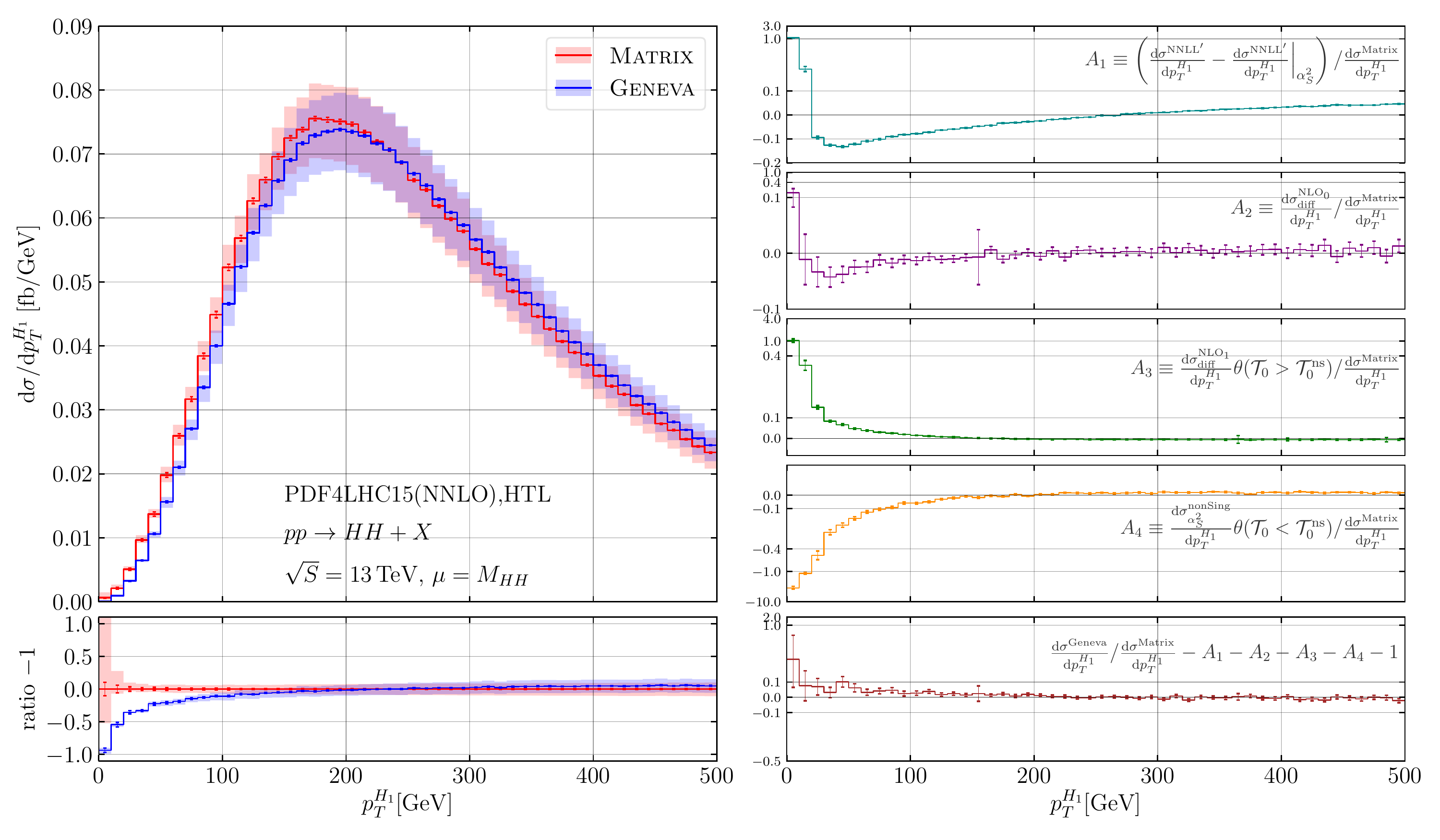}
  \end{center}
  \spaceabovefigurecaption
  \caption{\label{fig:higgs1pt_matrix}Comparison of the transverse momentum of the
    hardest Higgs boson between \Matrix and
    \geneva (left) and breakdown of all relevant contributions that account for the differences.
    (right).
  }
  \spacebelowfigurecaption
\end{figure}

To see how such differences can arise between results that are formally at the
same accuracy, we studied the discrepancies for $p_T$ of $H_1$ and $H_2$ and the
$\chi$ distribution, and in the following we report detailed results for the
transverse momentum distribution of the hardest Higgs boson, where these effects
are largest.
In Fig.~\ref{fig:higgs1pt_matrix}, we report a breakdown of all possible
sources of differences between a purely FO calculation such as that
obtained via \Matrix and the \geneva partonic prediction. All quantities shown in the sub-panels are
expressed in terms of their relative size to the total distribution obtained with
\Matrix, and we detail each of them in the following.
Firstly, it can be seen that in the small transverse momentum region we have a
discrepancy --  considerably bigger in size than that observed with the $p_T$ of the softest
Higgs boson -- which can reach  $100 \%$.
In the upper right panel of Fig. \ref{fig:higgs1pt_matrix} the first term that we look at is $A_1$, defined as
\begin{equation}
  A_1 \equiv \frac{ \frac{\de \sigma^{\rm NNLL^\prime}}{\de p_T^{H_1}} - \frac{\de \sigma^{\rm NNLL^\prime}}{\de p_T^{H_1}}\bigg|_{\alpha_S^2} }{\frac{\de \sigma^{\Matrix}}{\de p_T^{H_1}}},
\end{equation}
which represents the difference between the resummed contribution and the resummed
expanded up to $\ord{\alpha_S^2}$ coming from \geneva, normalised to the \Matrix
NNLO result. This difference is purely due to logarithmic terms beyond NNLO, and,
as can be
seen, in the region of interest gives rise to a large positive effect.
In the second and third ratio plot we consider the contributions from
projectable configurations with $\Tau_0 < \TauZcutns$ at relative $\mathcal{O}(\alpha_S)$, $A_2$,  and with $\Tau_0 > \TauZcutns$ and $\Tau_1 < \Tau_1^{\cut}$ at relative $\mathcal{O}(\alpha_S^2)$, $A_3$, defined as
\begin{align}
  A_2 \equiv \frac{ \frac{\de \sigma^{\rm NLO_0}_{\it diff}}{\de p_T^{H_1}} }{\frac{\de \sigma^{\Matrix}}{\de p_T^{H_1}}}, \qquad A_3 \equiv \frac{\frac{\de \sigma^{\rm NLO_1}_{\it diff}}{\de p_T^{H_1}} \theta(\Tau_0 > \TauZcutns)}{\frac{\de \sigma^{\Matrix}}{\de p_T^{H_1}}}\,,
\end{align}
where the subscript \textit{diff} refers to taking the difference
between the observables evaluated on exact kinematical configurations, and those
evaluated on projected kinematics below the respective resolution cutoffs.
The effect of these two terms are pure power corrections as they
arise from the projections used in \geneva to assign the event
kinematics below the resolution cutoffs.
Once again they are large and positive.
Lastly, we examine the following quantity,
\begin{equation}
  A_4 \equiv \frac{\frac{\de \sigma^{\rm nonSing}_{\alpha_S^2}}{\de p_T^{H_1}} \theta(\Tau_0 < \TauZcutns)}{\frac{\de \sigma^{\Matrix}}{\de p_T^{H_1}}},
\end{equation}
which represents the difference between the pure $\ord{\alpha_S^2}$
contributions of \geneva and \Matrix below $\TauZcutns$, normalised to the
same value as before. This last term corresponds to the contribution
in Eq.~\eqref{eq:nonsingcumas2}, projected onto
$p_T^{H_1}$. Note that, in order to compute this quantity, one needs to take the
$\mathcal{O}(\alpha_S^2)$ from \Matrix and subtract it from the
resummed-expanded \geneva result at the same order.
As shown in the respective sub-panel, this difference is negative and much larger
than those considered above, thus being the main driver of the discrepancy.
To conclude, and to show that there are no other possible sources of differences between \geneva and \Matrix,  we report in the last panel the difference between the partonic
\geneva result and \Matrix, subtracted of all the $A_i$ contributions which, as expected, is compatible
with zero.

%%%%%%%%%%%%%%%%%%%%%%%%%%%%%%%%%%%%%%%%%%%%%%%%%%%%%%%%%%%%%%%%%%%%%%%%%%%%%%%%
\section{Parton Shower Interface}
\label{sec:shower}
%%%%%%%%%%%%%%%%%%%%%%%%%%%%%%%%%%%%%%%%%%%%%%%%%%%%%%%%%%%%%%%%%%%%%%%%%%%%%%%%
The general idea behind how the interface to the parton shower is performed in
\geneva has been presented in various references, see in particular
Ref.~\cite{Alioli:2015toa} for a detailed discussion. As such, in this context,
we limit to a brief recap of the main features, highlighting the main novelties
introduced for this specific process.
The main issues one faces when matching a resummed calculation to the parton
shower is to ensure that both the accuracy of the variable in which the
resummation is performed and the accuracy of the parton shower are preserved.
\begin{figure}[t]
  \begin{center}
    \includegraphics[width=\rescaleoneplot]{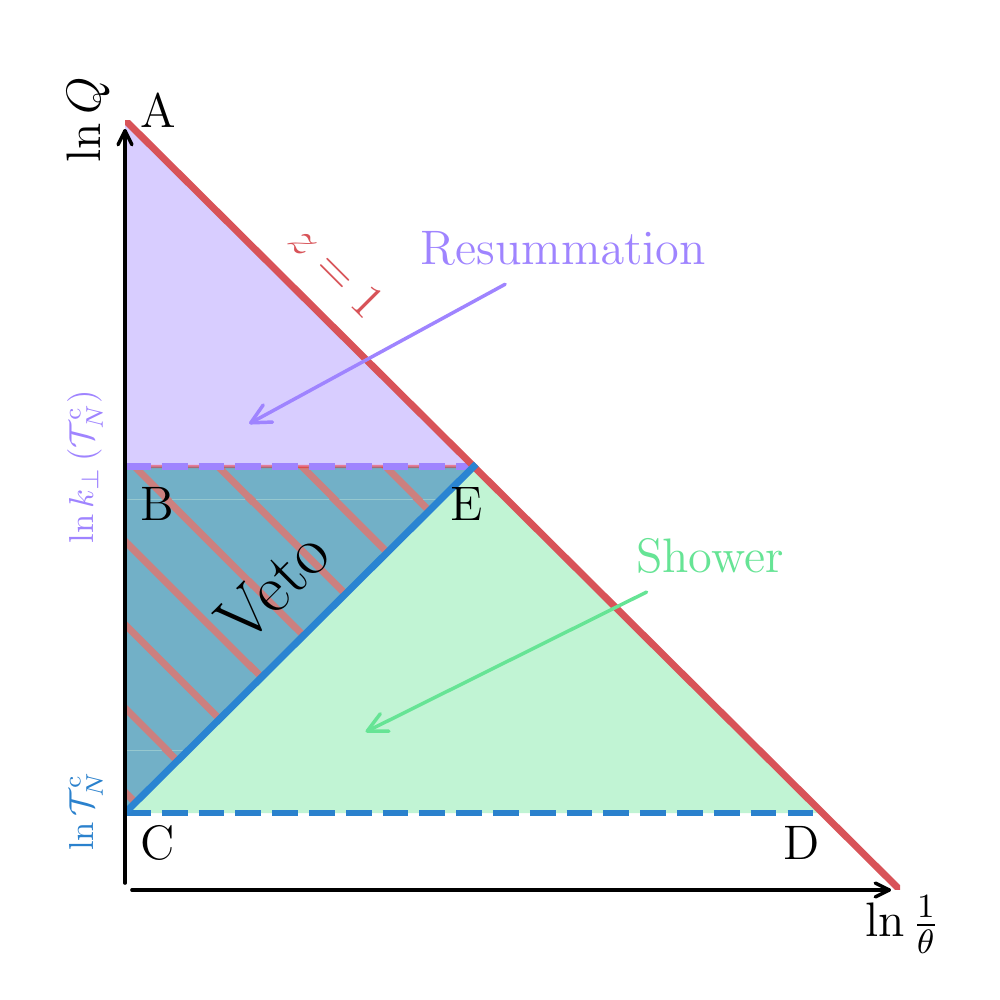}
    \caption{\label{fig:lundplane} Lund Plane representation of shower emissions and their interplay
      with the resummation region. }
  \end{center}
\end{figure}

In the specific case at hand, the resummation is performed in $\mathcal{T}_N$,
from a hard scale $Q$ down to a lower scale $\mathcal{T}_N^{c}$. This can
be
represented graphically, as done in Fig.~\ref{fig:lundplane} with the purple-shaded
triangular region (ACE), in the phase-space that an extra emission can have (commonly
referred to as Lund Plane).
As explained in the previous sections, \geneva
produces events with $0,1$ or $2$ final-state partonic jets, each of which determines
different lower scales for the resummation, corresponding to
$\TauZcutre$, $\mathcal{T}_1^{\mathrm{cut}}$ and
$\mathcal{T}_1\left(\Phi_2\right)$, respectively. The value of this scale
corresponds to a diagonal line on the Lund Plane, and its intersection with the
maximum available energy fraction an emission can have (E)  sets the maximum relative
transverse momentum of an emission ($k_\perp(\mathcal{T}_N^{c})$) given the
lower resummation scale ($\Tau_N^{c}$). This, in turn, determines the starting scale for the parton
shower. Now, the parton showers considered in this work produce emissions
ordered in the relative transverse momentum, meaning that the allowed region for
any emission from the parton shower is given by the trapezium (BCDE) spanning from the
horizontal line determined by $k_\perp(\mathcal{T}_N^{c})$ to that determined by
$\mathcal{T}_N^{c}$.
This clearly produces a double-counted region, identified by the hashed
triangle (BCE), where the shower could in principle produce emissions in the
resummation region. To avoid this, we perform a veto procedure -- much like what
is done in matrix element merging techniques, such as
CKKW-L~\cite{Lonnblad:2001iq} -- that consists in discarding and retrying any event for which,
after showering, we obtain a value of $\mathcal{T}_N > \mathcal{T}_N^{c}$.

To see that this procedure correctly ensures that no single shower emission can end up
in the vetoed region, thus spoiling the NNLL
accuracy of our $\Tau_0$ spectrum, recall that for any given
final-state multiplicity $M$, $\mathcal{T}_M(\Phi_M) = 0$. Thus, for any one emission
from the parton shower that produces a final state with $M+1$ partons, we have
that $\mathcal{T}_M(\Phi_{M+1}) = \mathcal{T}$ encodes the hardness of that
emission, and can be compared to $\mathcal{T}_N^{c}$. In addition, $N$-jettiness
is an additive variable of strictly positive terms, and the following relation,
for a given final-state multiplicity, holds:
\begin{equation}
  \mathcal{T}_{N}(\Phi_M) \geq \mathcal{T}_{N+1}(\Phi_M)\, .
\end{equation}
To prove then that no single emission enters the vetoed region, consider the
case the parton shower starts with a $N$-parton configuration, with
$N=0,1,2$. After the parton shower, we end up with a $N+k$ partonic final state, where
$k$ stands for the total emissions performed by the shower. Our veto implies
that
\begin{equation}
  \mathcal{T}_{N} (\Phi_{N+k}) \leq  \mathcal{T}_{N}^{c}\,.
\end{equation}
Combining the previous relations, we get that
\begin{equation}
  \label{eq:chainofinequalities}
  \mathcal{T}_{N+k-1} (\Phi_{N+k})\leq \mathcal{T}_{N+k-2} (\Phi_{N+k}) \leq
  \dots\leq\mathcal{T}_{N} (\Phi_{N+k}) \leq  \mathcal{T}_{N}^{c}\,,
\end{equation}
which implies that also the hardness of each of the $k$th emissions has
to be smaller than $\mathcal{T}_{N}^{c}$, given the additive property of $N$-jettiness.
This means that the final events accepted after the veto could always be generated via a sequence
of $\Tau$-ordered emissions, even if the actual showering might not have respected that
condition locally.
To better clarify the meaning of Eq.~\eqref{eq:chainofinequalities}, let us
consider the following explicit example. Imagine that we start the parton shower
from a configuration with two partons, $\Phi_2$, which represents the bulk of the events generated
by \geneva after the resummation of $\Tau_0$ and $\Tau_1$ has been performed at the parton level.
Thus, the value of
$\mathcal{T}_N^{c} = \mathcal{T}_1(\Phi_2)$ determines the starting scale of the
parton shower $k_\perp(\mathcal{T}_1(\Phi_2))$. An additional emission of the
shower has a hardness of $\Tau = \mathcal{T}_2(\Phi_3)$ which by construction
has to be smaller than $\mathcal{T}_N^{c}$ to respect the veto condition. When a
second emission is performed its hardness is given by
$\Tau = \mathcal{T}_3(\Phi_4) \leq \mathcal{T}_2(\Phi_4)$ which by construction
needs to be smaller than $\mathcal{T}_1(\Phi_2)$ to satisfy the veto. Iterating
this, one can reconstruct the full chain of inequalities appearing in
Eq.~\eqref{eq:chainofinequalities}.
As the variable $N$-jettiness is additive, it is also true that the
hardness of the first emission is constrained by the veto to be lower
than the lower resummation scale.
Clearly this does not impose any ordering
between the hardness of the two emissions, $\Tau_2(\Phi_3)$ and $\Tau_3(\Phi_4)$, it just
implies that they are both lower than $\mathcal{T}_1(\Phi_2)$.
Note that, as a consequence, this implies that the accuracy of the parton
shower is not spoiled by our matching procedure. In fact, the resummation region
has a higher accuracy than that of the shower, and in the shower region there is
no double-counted contribution. Thus the accuracy on any observable computed
with this matching is at least as accurate as the parton shower is.
\begin{figure}[t]
  \begin{center}
    \includegraphics[width=\rescaleoneplot]{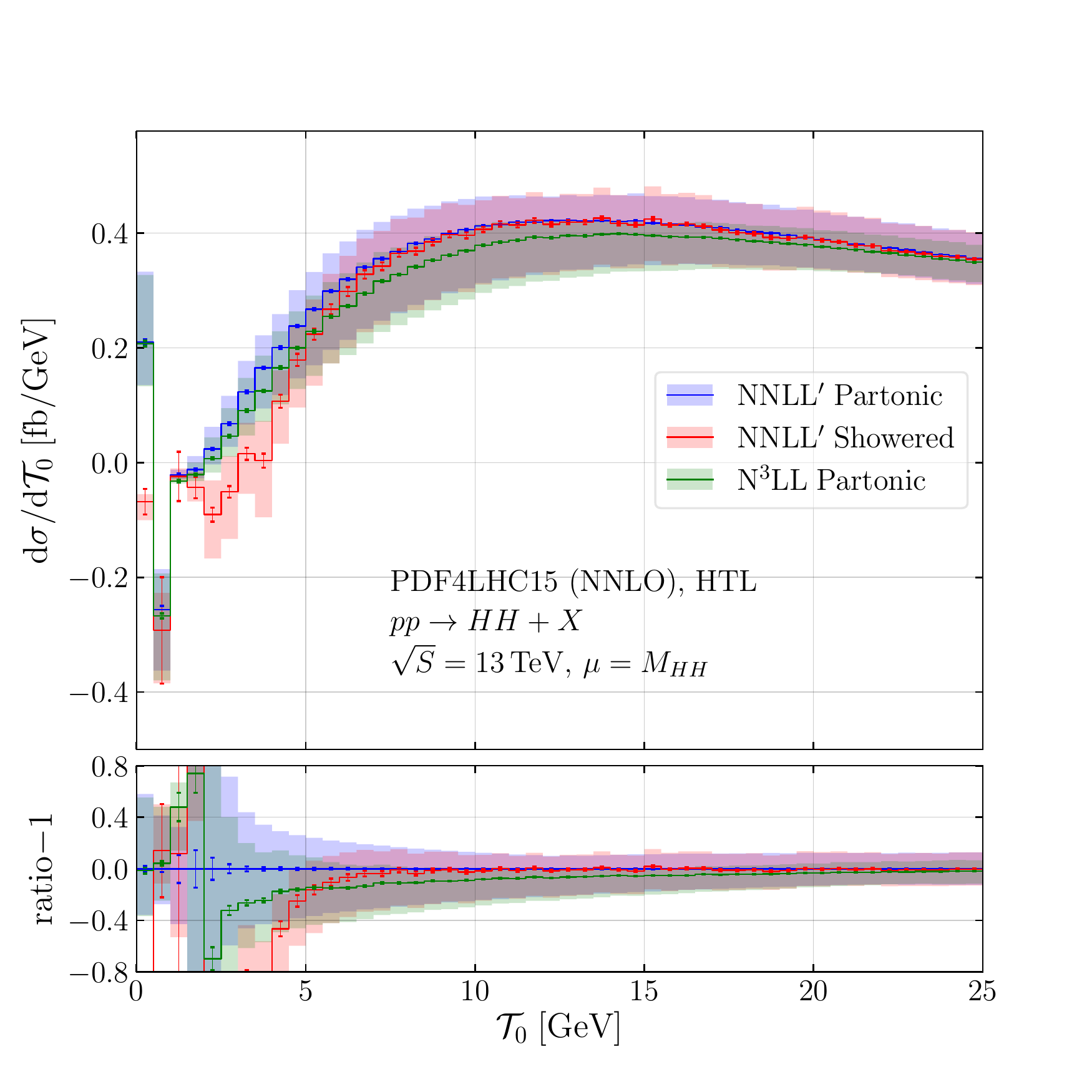}
    \spaceabovefigurecaption
    \caption{\label{fig:showervsresummation} Comparison of the effects of the
      shower and the effects of the N$^3$LL resummation.}
    \spacebelowfigurecaption
  \end{center}
\end{figure}

The argument presented above does not imply that the resummed variable --
$\mathcal{T}_0$ in this case -- is numerically preserved, as shown in
Fig.~\ref{fig:showervsresummation}. However, it can be seen that the shift in
the spectrum induced by the parton shower is of similar size as that given by
including higher-order effects in the resummation. Compared to other
colour-singlet production processes studied in the past with the \geneva
framework, one can notice that, in this case, the impact of the shower on the $\mathcal{T}_0$
distribution can be numerically sizeable. This has to do with two key
differences with respect to other
processes~\cite{Alioli:2021egp,Alioli:2019qzz,Cridge:2021hfr,Alioli:2021qbf}.
First, since this process is dominated
by gluon channels, we expect effects due to gluon emissions to be scaled
by a factor $C_A/C_F \sim 2$.  Second, this is one of the first processes which
features a hard scale, $M_{HH}$, which spans different orders of magnitude and
does not really present a sharp peak. This implies that even at relatively large
values of $\mathcal{T}_0$ -- given that the scale we have control over is
$\tau_0=\mathcal{T}_0/M_{HH}$ -- one can have a small value of $\tau_0$ with
$M_{HH}$ still large. The consequence is that, for any fixed value of $\mathcal{T}_0$,
the large logarithmic terms associated to this Higgs pair production can be
significantly larger than the corresponding terms for the same value of $\Tau_0$
in other processes.

\begin{figure}[t]
  \begin{center}
    \includegraphics[width=\rescaletwoplots]{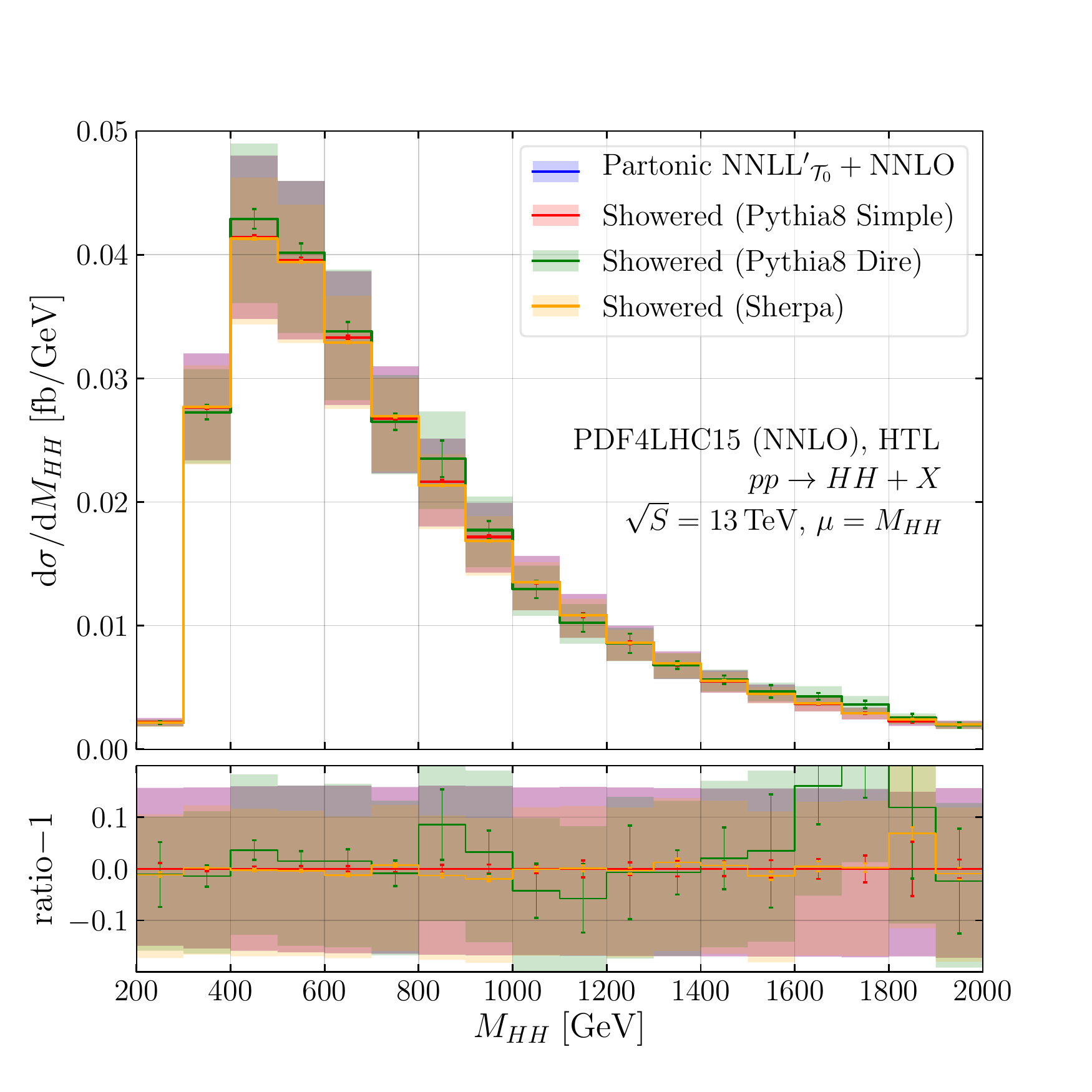}
    \includegraphics[width=\rescaletwoplots]{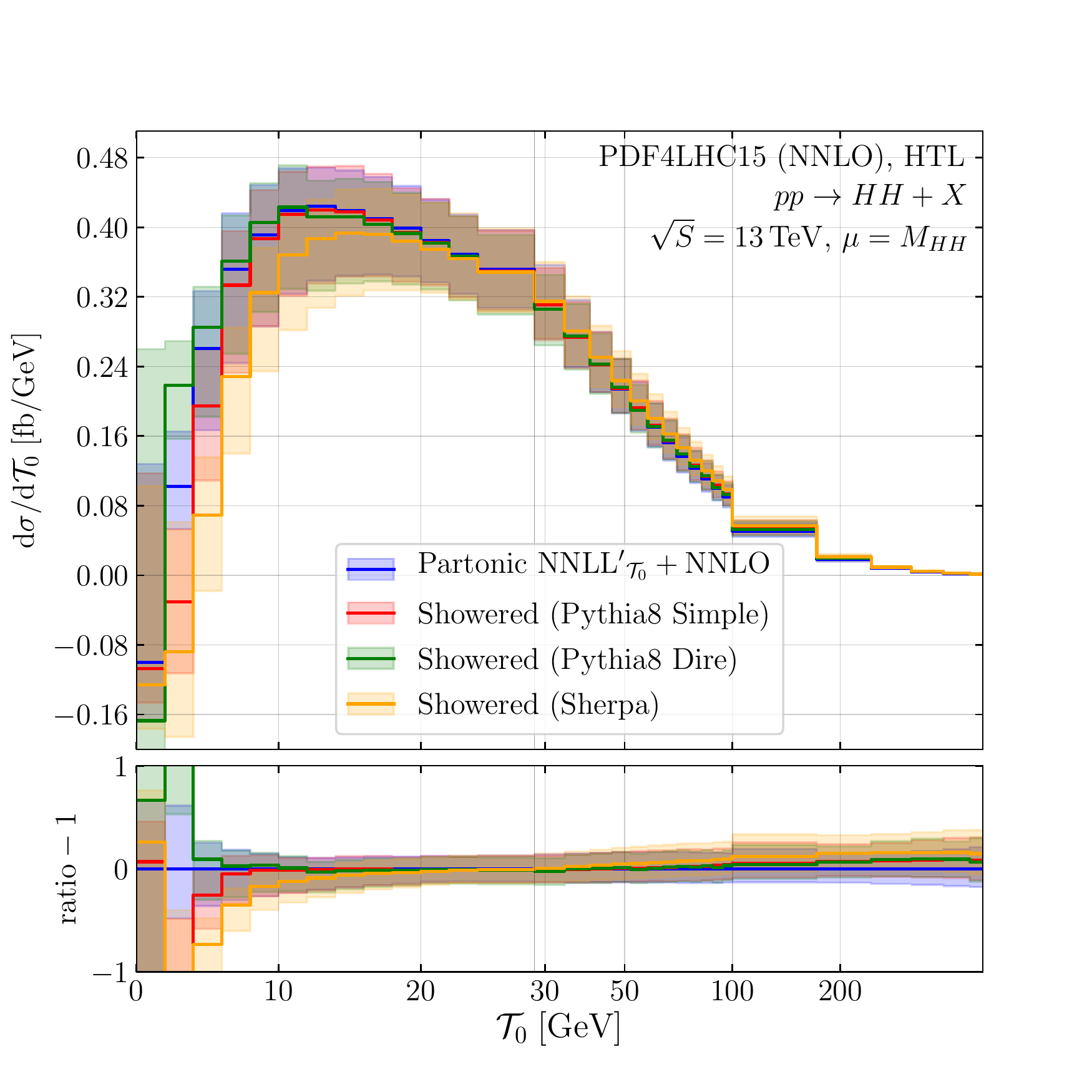}
    \spaceabovefigurecaption
    \caption{\label{fig:tau0showercomparison}Comparison between the partonic and
      showered for the di-Higgs invariant-mass
      (left panel) and the $\Tau_0$ (right panel) distributions in \geneva,
      \geneva+~\pythia8, \geneva+~\dire and \geneva+~\sherpa.}
    \spacebelowfigurecaption
  \end{center}
\end{figure}

In addition, as a validation of the matching with the shower, we show in the
left panel of
Fig.~\ref{fig:tau0showercomparison} how the shower correctly preserves the
spectrum of fully inclusive variables such as the invariant mass of the Higgs
pair. This implies that the total inclusive cross section is also preserved by
the shower.

To further study the impact of parton showers, we extend \geneva's default
shower interface to \pythia8 to both \dire~\cite{Hoche:2015sya}, as implemented
in \pythia8, and the default shower in
\sherpa~\cite{Gleisberg:2008ta,Sherpa:2019gpd,Schumann:2007mg}. These three
parton showers differ most notably by the choice of the evolution variable, which,
together with the starting scale imposed by our matching, determines how much of
the phase space away from the strict soft and collinear limits is available to the
parton shower.

As shown in Fig.~\ref{fig:tau0showercomparison}, right panel,
the impact of the choice of the parton shower can be relatively
large for small ($\leq 10$~GeV) values of $\mathcal{T}_0$.
Indeed, while all the three showers present deviations from the partonic result
of roughly the same magnitude, they differ significantly among each other, which
can be approximately viewed as a shower matching uncertainty. The fact that this
process can be highly sensitive to different choices of evolution variables, and
thus different parton showers, was shown in Ref.~\cite{Jones:2017giv}. The net result
is that, for the default choice of the evolution variable, the Catani-Seymour based
shower as implemented in \sherpa has an evolution variable $t$ on average
larger than that of \dire, leading to its phase space reach being more
constrained. This is reflected in the suppression in the small $\mathcal{T}_0$
region.

\begin{figure}[tp]
  \begin{center}
    \includegraphics[width=\rescaletwoplots]{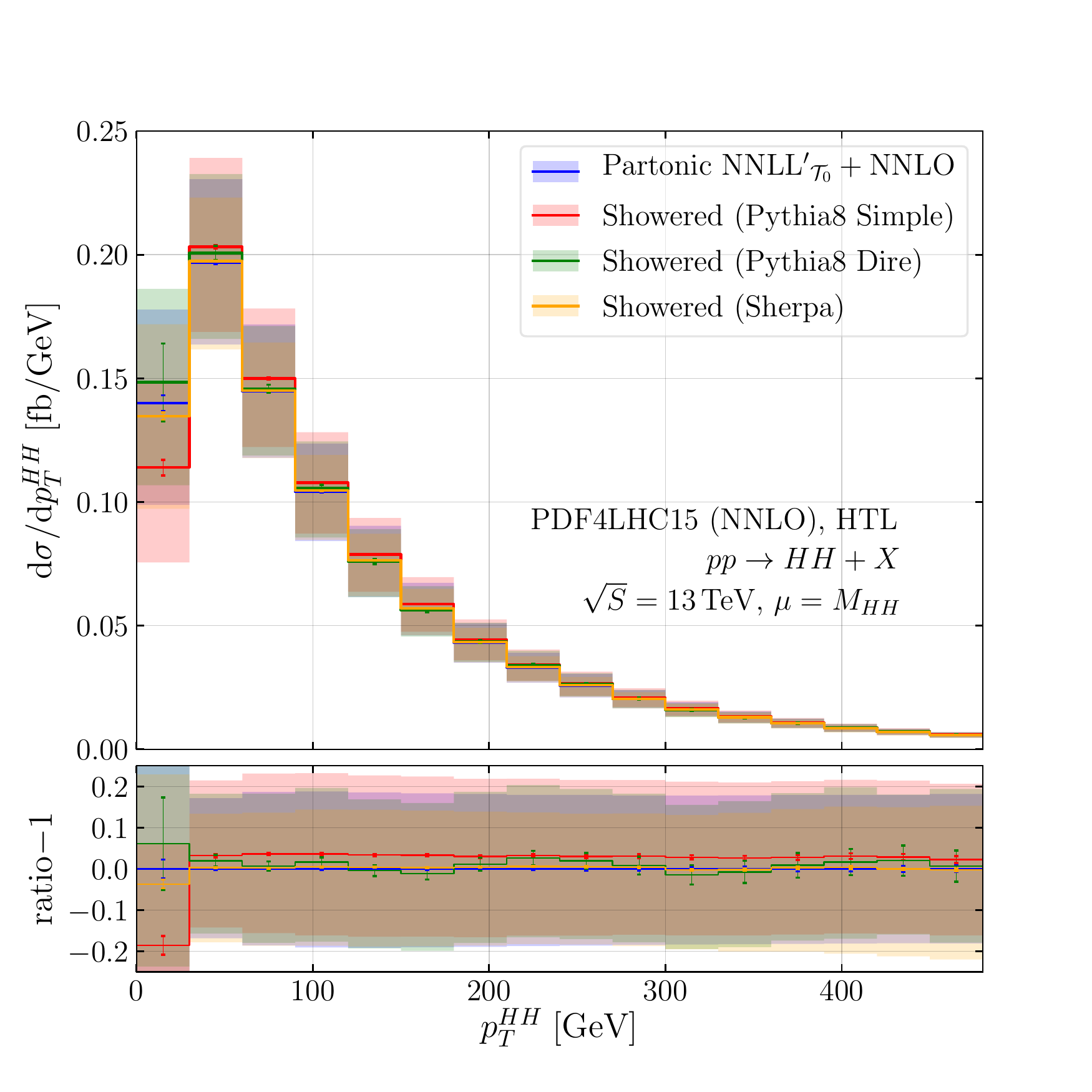}
    \includegraphics[width=\rescaletwoplots]{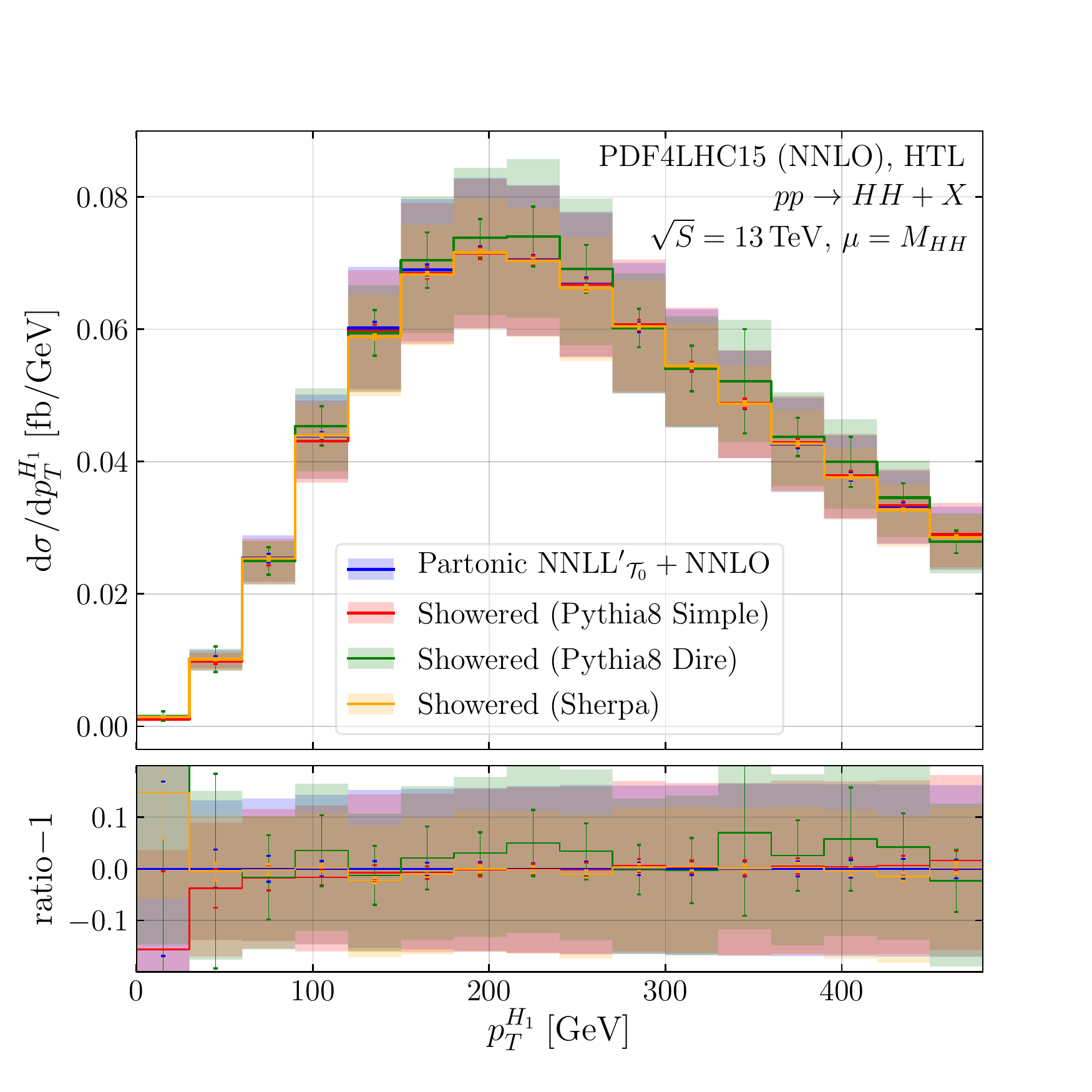}
  \end{center}
  \spaceabovefigurecaption
  \caption{\label{fig:pthhandpth1}Comparison between the partonic and showered for
    $p_T^{HH}$ (left) and $p_T^{H_1}$ (right) in \geneva,
    \geneva+~\pythia8, \geneva+~\dire and \geneva+~\sherpa.}
  \spacebelowfigurecaption
\end{figure}

Lastly, in Fig.~\ref{fig:pthhandpth1} we show how the different showers affect
the transverse momentum of the Higgs pair system and that of the hardest Higgs boson.
In this case, although in principle the parton shower is not required to
preserve either of these observables, we see that all shower predictions largely
agree among themselves and with the partonic result, aside from the very first
bin where they agree within uncertainties.
This is likely due to the fact that partonic events produced by \geneva already
feature both a $\mathcal{T}_0$ and a $\mathcal{T}_1$ resummation,
both of which have a non-trivial interplay with the transverse momentum
distribution of the colour-singlet system.

%%%%%%%%%%%%%%%%%%%%%%%%%%%%%%%%%%%%%%%%%%%%%%%%%%%%%%%%%%%%%%%%%%%%%%%%%%%%%%%%
\section{Conclusions}
\label{sec:conclusions}
%%%%%%%%%%%%%%%%%%%%%%%%%%%%%%%%%%%%%%%%%%%%%%%%%%%%%%%%%%%%%%%%%%%%%%%%%%%%%%%%

With the detailed study of the properties of the Higgs boson, discovered ten
years ago at the LHC, set to dominate the research focus of the next twenty to fifty
years, the ability to constrain the self-coupling of the Higgs boson is one of the
fundamental milestones. While many of such properties are largely dominated by
the reach of the experimental set-up, the Higgs self-coupling determination is
instead limited by statistics, from the experimental point of view, and by
uncertainties in the theoretical predictions for both signal and
background. Indeed, even after the high-luminosity phase of the LHC, it is predicted
that with the current theoretical knowledge the value of the self-coupling will
only be constrained to about $50\%$~\cite{DiMicco:2019ngk}.

From the theoretical standpoint, the process affecting the most the
determination of this fundamental parameter of the Standard Model Lagrangian, at
hadron colliders, is
the production of a Higgs pair in the gluon-gluon fusion channel. Similarly to the
production of a single Higgs boson this process proceeds via a top-quark loop, thus
rendering the inclusion of higher corrections in the exact theory, with full heavy-quark mass dependence, highly
non-trivial. Indeed only NLO corrections are known exactly~\cite{Borowka:2016ehy} for
this process, as they require a state-of-the-art two-loop calculation with both internal and external masses.
Nevertheless, similarly to the single Higgs case, one can expand the
Lagrangian of the Standard Model in inverse powers of the top-quark mass thus
constructing an effective theory where the top quark is decoupled and Higgs bosons
couple directly to gluons, usually referred to as HTL.
While this approach is known to be a poor approximation for double Higgs production, it
still provides useful insights if the main interest is to study the effects of
the inclusion of higher-order QCD corrections, resummation and the interface to the  parton shower.

In this work we take the latter approach and present an implementation of the
production of a pair of Higgs bosons via gluon-gluon fusion in the HTL in
\geneva. By employing this framework we are able to produce
fully-differential NNLO results matched to a $\Tau_0$ NNLL$^\prime$ resummation, using
the SCET formalism, further
interfaced to the parton shower. We show that, using a traditional value of the
hard scale in the argument of the logarithms, such as the Higgs pair
invariant mass, resummation effects
affect exclusive observables at much larger values
than in processes previously studied.
This is a consequence of the fact that the $M_{HH}$ distribution spans over a large range.
Moreover, we study in detail how the inclusion of resummation and subleading
power corrections can impact the comparison with standard fixed-order results,
such as those obtained with \Matrix, tracking down all sources of possible differences at higher order.
Lastly, we present and discuss the impact of
interfacing our partonic predictions to different parton showers and find that this process is subject to large parton shower
effects.
This is the first step towards a more comprehensive estimation of parton shower
matching uncertainties.

It is clear that, while the approach taken in this work to use the HTL works
well as far as we only discuss the effect of pure QCD higher-order effects, in order to
develop a realistic event generator for the production of a Higgs pair, we need
to include mass effects if not in an exact way, at least in some approximation. Indeed,
the natural continuation of this work is to explore ways, such as those devised
in Ref.~\cite{Davies:2019nhm}, to include partial top-quark mass effects in our
implementation.
%%%%%%%%%%%%%%%%%%%%%%%%%%%%%%%%%%%%%%%%%%%%%%%%%%%%%%%%%%%%%%%%%%%%%%%%%%%%%%%%
\section*{Acknowledgements}
\label{sec:Acknowledgements}
%%%%%%%%%%%%%%%%%%%%%%%%%%%%%%%%%%%%%%%%%%%%%%%%%%%%%%%%%%%%%%%%%%%%%%%%%%%%%%%%

We thank F.~Tackmann for comments on the manuscript and for providing us with a
preliminary version of the \texttt{scetlib} library. We are also grateful to
S.~H\"oche for useful discussions on the shower accuracy.
This project has received funding from the European Research Council (ERC)
under the European Union's Horizon 2020 research and innovation programme
(Grant agreements No. 714788 REINVENT and 101002090 COLORFREE)
The work of GB is supported by the FARE grant R18ZRBEAFC.
SA also acknowledges funding from Fondazione Cariplo and Regione
Lombardia, grant 2017-2070 and from MIUR
through the FARE grant R18ZRBEAFC. MAL is supported by the Deutsche
Forschungsgemeinschaft (DFG) under Germany's Excellence Strategy -- EXC 2121 ``Quantum Universe''
-- 390833306. We acknowledge the CINECA and the National Energy Research Scientific Computing Center (NERSC), a U.S. Department of Energy Office of
Science User Facility operated under Contract No. DEAC02-05CH11231,
for the availability of the high performance computing resources
needed for this work.

\appendix

\section{Hard function for $gg \to HH$ in the $\overline{\rm MS}$  scheme}
\label{appendix_A}
In this appendix, we report the necessary ingredients to obtain the hard
function in the $\overline{\mathrm{MS}}$ scheme.
Its perturbative expansion, in a generic subtraction scheme $X$ reads
\begin{equation}\label{eq:hard_function}
  H_X = \left( \frac{\alpha_S}{4 \pi} \right)^2 \left[H^{(0)}+ \frac{\alpha_S}{4 \pi} H^{(1)}_X + \left( \frac{\alpha_S}{4 \pi} \right)^2 H^{(2)}_X + \ord{\alpha_S ^3}\right].
\end{equation}
Note that $H^{(0)}$ is scheme independent, since it is
given by the cross section at LO,
defined, for this process, as
\begin{align}
  H^{(0)} &= \frac{s \, C^2_{\rm LO}}{144 \, v^4}.
\end{align}
Defining $s$ as the squared centre-of-mass energy of the process, $v$ the vacuum expectation
value, $m_H$ the Higgs boson mass one has
\begin{align}
 C_{\rm LO} = \frac{6\lambda v^2}{s-m^2_H} -1,\qquad 2\lambda v^2 = m_H^2 \quad
  \mathrm{and} \quad v^4 = \frac{1}{2G_F^2}\,.
\end{align}
In the Catani subtraction scheme ($X=C$), the expressions for the one- and two-loop
coefficients of the hard function are given in Refs.~\cite{deFlorian:2013uza}
and~\cite{Grigo:2014jma}.
To extract the results needed for the implementation in \geneva, we start by defining
 \begin{equation}
   H_C^{(i)} = \frac{\de \sigma_{\mathrm{fin}}^{(i)}}{\de t}  \frac{\de t}{\de \Phi_2}
 \end{equation}
where we used the following Jacobian,
 \begin{equation}
   \frac{\de\Phi_2}{\de t} = \frac{1}{8\pi s}
   \frac{1}{\Gamma\<\L(1-\epsilon\R)} \L[\frac{s \L(s-4m_H^2\R) -
     \L(t-u\R)^2}{16\pi s}\R]^{-\epsilon}\,.
 \end{equation}
 We can express the coefficients of Eq.~\eqref{eq:hard_function} in the $\overline{\rm MS}$ scheme ($X=\overline{\rm MS}$)
 by exploiting the following relations
\begin{align}
  \label{eq:H12_msbar}
  H^{(1)}_{\overline{\rm MS}} &= H^{(1)}_C + \lim_{\epsilon \to 0} 2 \, {\rm Re} \, [ 2 \, \mathcal{I}^{(1)}(\epsilon) + \mathcal{Z}^{(1)}(\epsilon)] \, H^{(0)}, \nn \\
  H^{(2)}_{\overline{\rm MS}} &= H^{(2)}_C + \lim_{\epsilon \to 0} \bigg\{ 2 \, {\rm{Re}} \, [2 \, \mathcal{I}^{(1)}(\epsilon) + \mathcal{Z}^{(1)}(\epsilon)]\bigg\} \, H^{(1)}_C
  + \bigg\{ |\lim_{\epsilon \to 0} [ 2 \, \mathcal{I}^{(1)}(\epsilon) + \mathcal{Z}^{(1)}(\epsilon)] |^2 \nn \\
                                      &+ 2 {\rm{Re}} \lim_{\epsilon \to 0} \bigg( 4 \, \mathcal{I}^{(2)}(\epsilon) + 2 \, \mathcal{I}^{(1)}(\epsilon) [ 2 \, \mathcal{I}^{(1)}(\epsilon) + \mathcal{Z}^{(1)}(\epsilon) ] + \mathcal{Z}^{(2)}(\epsilon) \bigg) \bigg\} H^{(0)}\,,
\end{align}
where for simplicity we dropped the $\mu$ dependence from the $H^{(i)}_X$, $\mathcal{I}^{(i)}$ and $\mathcal{Z}^{(i)}$.
In the equation above, $\mathcal{I}^{(i)}$ are the perturbative coefficients
of the $\mathcal{I}$ operator defined as~\cite{deFlorian:2012za}
\begin{align}
  \mathcal{I}(\epsilon, \mu) = 1 + \left( \frac{\alpha_S}{2 \pi} \right) \mathcal{I}^{(1)}(\epsilon, \mu) + \left( \frac{\alpha_S}{2 \pi} \right)^2 \mathcal{I}^{(2)}(\epsilon, \mu) + \ord{\alpha_S^3} \, ,
\end{align}
where
\begin{align}
  \mathcal{I}^{(1)}(\epsilon, \mu) &= -\bigg(  \frac{\mu^2}{-s} \bigg)^{\epsilon} \frac{\exp(\epsilon \, \gamma_E)}{\Gamma(1-\epsilon) } \bigg( C_g \frac{1}{\epsilon^2} + \gamma_g \frac{1}{\epsilon}\bigg), \nn \\
  \mathcal{I}^{(2)}(\epsilon,\mu) &= \bigg(  \frac{ \mu^2}{-s} \bigg)^{\epsilon} \frac{\exp(\epsilon \gamma_E)}{72 \, \Gamma(1-\epsilon) \epsilon^4} \bigg\{ 12 \epsilon (C_g + \epsilon \gamma_g)(11C_A-2n_f) \nn \\
                                    &- \frac{36 \exp(\epsilon \gamma_E)}{(\Gamma(1-\epsilon)} \bigg( \frac{ \mu^2}{-s} \bigg)^{\epsilon} (C_g + \epsilon \gamma_g)^2\
                                      + \epsilon \bigg( \frac{\mu^2}{-s}
                                      \bigg)^{\epsilon}\bigg[ 36 \epsilon^2
                                      H_g\nn\\
   &+ 2 (3+5\epsilon)(C_g+2\epsilon \gamma_g )n_f + C_A(C_g+2\epsilon \gamma_g)(-33-67 \epsilon +3 \epsilon \pi^2) \bigg] \bigg\}\,,
\end{align}
and for gluon-initiated processes one finds
\begin{align}
  C_g &= C_A, \qquad \gamma_g = \frac{ \beta_0}{2}, \nn \\
  H_g &= C^2_A \bigg( \frac{1}{2} \zeta_3 + \frac{5}{12} + \frac{11 \pi^2}{144} \bigg) - C_A n_f \bigg( \frac{29}{27} + \frac{\pi^2}{72}\bigg) + \frac{1}{2} C_F n_f
        + \frac{5}{27} n^2_f\,.
\end{align}
The $\mathcal{Z}^{(i)}$ in Eq.~\eqref{eq:hard_function} are obtained
from the following expansion of the $\mathcal{Z}$ factor~\cite{Becher:2009qa},
\begin{align}
 \mathcal{Z}^{-1}(\epsilon, \mu) =	1 + \left( \frac{\alpha_S}{4 \pi} \right) \mathcal{Z}^{(1)}(\epsilon, \mu) + \left( \frac{\alpha_S}{4 \pi} \right)^2 \mathcal{Z}^{(2)}(\epsilon, \mu) + \ord{\alpha_S^3} \, ,
\end{align}
where
\begin{align}
 \mathcal{Z}^{(1)}(\epsilon,\mu)&= - \frac{\Gamma'_0}{4 \epsilon^2} - \frac{\mathbf{\Gamma}_0}{2\epsilon}, \nn \\
 \mathcal{Z}^{(2)}(\epsilon,\mu) &=  \frac{(\Gamma'_0)^2}{32 \epsilon^4} + \frac{3 \beta_0 \Gamma'_0+2\Gamma'_0 \mathbf{\Gamma}_0}{16 \epsilon^3} + \frac{4 \beta_0 \mathbf{\Gamma}_0 + 2\mathbf{\Gamma}_0^2- \Gamma'_1}{16 \epsilon^2} - \frac{\mathbf{\Gamma}_1}{4 \epsilon} \,,
\end{align}
and
\begin{align}
    \mathbf{\Gamma}_i &= -C_A \Gamma_i \ln \bigg( \frac{\mu^2}{-s} \bigg)+ 2
               \gamma^g_i, \qquad \Gamma'_i = -2 C_A \Gamma_i \,,\nn\\
  \gamma^g_1 &= C^2_A \bigg( - \frac{692}{27} + \frac{11 \pi^2}{18} + 2 \zeta_3 \bigg) + C_A T_F n_f \bigg( \frac{256}{27} - \frac{2 \pi^2}{9} \bigg)
  + 4 C_F T_F n_f \, ,
\end{align}
with $C_A$, $C_F$ the colour factors, $n_f$ the number of light flavours and $T_F = 1/2$.
Considering Eq.~\eqref{eq:H12_msbar} and setting $\mu^2=s$, we find the following results for the
translation to the $\overline{\rm MS}$ scheme of the hard function coefficients
\begin{align}
 H^{(1)}_{\overline{\rm MS}} &= H^{(1)}_C + \frac{7 C_A \pi^2}{3} H^{(0)}, \nn \\
 H^{(2)}_{\overline{\rm MS}} &= H^{(2)}_C + \frac{7 C_A \pi^2}{3} H^{(1)}_C \nn \\
 &+ \bigg( \frac{167}{6}C^2_A \pi^2 - \frac{367}{54}C_A n_f \pi^2 + \frac{11 n^2_f \pi^2}{27}
  + \frac{73}{36}C^2_A \pi^4 - \frac{11}{3}C^2_A \zeta_3 + \frac{2}{3}C_A n_f \zeta_3 \bigg) H^{(0)}.
\end{align}
Finally, in order to restore the exact $\mu$ dependence of the hard function we
use the RGE equation
\begin{align}
  \label{eq:rge}
  \frac{\de}{\de \ln(\mu^2)} H(\mu^2) =  {\rm{Re}}[\mathbf{\Gamma}(\mu^2) ] \, H(\mu^2) .
\end{align}
Taking the first order of the expansion $\ord{\alpha_S^3}$, we have:
\begin{align}
  \frac{\de}{\de\ln(\mu^2)} H^{(1)}(\mu^2) -2 \beta_0 H^{(0)} &=  {\rm Re} [\mathbf{\Gamma}_0 (\mu^2)] H^{(0)} \, ,
\end{align}
having used
\begin{align}
  \frac{1}{4 \pi}\frac{\de}{\de \ln (\mu^2)} \alpha_S = - \bigg( \frac{\alpha_S}{4 \pi} \bigg)^2 \sum_{n=0} \bigg( \frac{\alpha_S}{4 \pi} \bigg)^n \beta_n.
\end{align}
For the second order of the expansion $\ord{\alpha_S^4}$ in Eq. \eqref{eq:rge} we have
\begin{align}
  \frac{\de}{\de \ln (\mu^2)}H^{(2)}(\mu^2) - 2 \beta_1 H^{(0)} -3 \beta_0 H^{(1)}(\mu^2) =  {\rm Re}[\mathbf{\Gamma}_0 (\mu^2)]H^{(1)}(\mu^2) +  {\rm Re}[\mathbf{\Gamma}_1(\mu^2)] H^{(0)}\,,
\end{align}
where $\beta_1$ is defined in Eq.~\eqref{eq:cuspnoncuspad}.
In conclusion, the hard function in the $\overline{\rm MS}$ scheme is given by
\begin{align}
  H_{\overline{\rm MS}}(\mu^2) =\left( \frac{\alpha_S}{4 \pi} \right)^2 \left[ H^{(0)}(\mu^2) + \left( \frac{\alpha_S}{4 \pi} \right) H^{(1)}_{\overline{\rm MS}}(\mu^2) + \left( \frac{\alpha_S}{4 \pi} \right)^2 H^{(2)}_{\overline{\rm MS}}(\mu^2) + \ord{\alpha_S^3}\right] \, ,
\end{align}
where
\begin{align}
  H^{(0)}(\mu^2) &=  H^{(0)}, \nn \\
  H^{(1)}_{\overline{\rm MS}}(\mu^2) &= H^{(1)}_{\overline{\rm MS}}(s) -  2 C_A \ln^2 \left(  \frac{\mu^2}{s} \right) H^{(0)}, \nn \\
  H^{(2)}_{\overline{\rm MS}}(\mu^2) &= H^{(2)}_{\overline{\rm MS}}(s) -H^{(1)}_{\overline{\rm MS}}(s) \left[ 2 C_A \ln^2 \left( \frac{\mu^2}{s}\right) - \frac{11}{3} C_A \ln \left( \frac{\mu^2}{s} \right) + \frac{2}{3}n_f \ln \left( \frac{\mu^2}{s} \right) \right] \nn \\
                 & + H^{(0)} \bigg[ \ln \left( \frac{\mu^2}{s} \right) \left( - \frac{772}{27}C^2_A + \frac{76}{27} C_A n_f + \frac{11}{9}C^2_A \pi^2 - \frac{2}{9} C_A n_f \pi^2 + 4 C^2_A \zeta_3 \right) \nn \\
                 &+ \ln^2 \left( \frac{\mu^2}{s} \right) \left( -\frac{134}{9}C^2_A + \frac{20}{9}C_A n_f + \frac{2}{3}C^2_A\pi^2 \right)
                   + \ln^3 \left( \frac{\mu^2}{s} \right) \left(- \frac{22}{9} C^2_A + \frac{4}{9} C_A n_f \right) \nn \\
                 &+ 2 C^2_A \ln^4 \left( \frac{\mu^2}{s} \right) \bigg].
\end{align}

\newpage
%% Bibliography
\bibliographystyle{JHEP}
\bibliography{geneva}

\end{document}